\documentclass [twocolumn]{aastex62}

\received{May 1, 2018}
\revised{May 7, 2018}
\accepted{\today}
\submitjournal{ApJ}

\shorttitle{Intragroup light in NGC 5018}
\shortauthors{Spavone et al.}

\begin{document}

\title{VEGAS: A VST Early-type GAlaxy Survey.\\III. Mapping the galaxy structure, interactions and intragroup light in the NGC 5018 group}

\correspondingauthor{Marilena Spavone}
\email{marilena.spavone@oacn.inaf.it}

\author{Marilena Spavone}
\affiliation{INAF-Astronomical Observatory of Capodimonte\\
 Salita Moiariello 16, 80131, Naples, Italy}
\nocollaboration

\author{Enrichetta Iodice}
\affiliation{INAF-Astronomical Observatory of Capodimonte\\
 Salita Moiariello 16, 80131, Naples, Italy}
\nocollaboration

\author{Massimo Capaccioli}
\affiliation{University of Naples Federico II, C.U. Monte Sant'Angelo\\
             via Cinthia, 80126, Naples, Italy}
\nocollaboration

\author{Daniela Bettoni}
\affiliation{INAF-Astronomical Observatory of Padova\\
 Vicolo dell'Osservatorio 5, I-35122, Padova, Italy}
\nocollaboration

\author{Roberto Rampazzo}
\affiliation{INAF-Astronomical Observatory of Padova\\
 Vicolo dell'Osservatorio 5, I-35122, Padova, Italy}
\nocollaboration

\author{Noah Brosch}
\affiliation{The Wise Observatory and School of Physics and Astronomy\\
             Tel Aviv University, Israel}
\nocollaboration

\author{Michele Cantiello}
\affiliation{INAF-Astronomical Observatory of Teramo\\ Via Maggini,
             64100, Teramo, Italy}
\nocollaboration

\author{Nicola R. Napolitano}
\affiliation{INAF-Astronomical Observatory of Capodimonte\\
 Salita Moiariello 16, 80131, Naples, Italy}
\nocollaboration

\author{Luca Limatola}
\affiliation{INAF-Astronomical Observatory of Capodimonte\\
 Salita Moiariello 16, 80131, Naples, Italy}
\nocollaboration

\author{Aniello Grado}
\affiliation{INAF-Astronomical Observatory of Capodimonte\\
 Salita Moiariello 16, 80131, Naples, Italy}
\nocollaboration

\author{Pietro Schipani}
\affiliation{INAF-Astronomical Observatory of Capodimonte\\
 Salita Moiariello 16, 80131, Naples, Italy}
\nocollaboration



\begin{abstract}

Most of the galaxies in the Universe at present day are in
  groups, which are key to understanding the galaxy evolution.
In this work we present a new deep mosaic of $1.2 \times 1.0$ square
 degrees of the group of galaxies centered on NGC~5018, acquired at the ESO VLT Survey Telescope.
We use  {\it u, g, r} images to analyse the structure of the group members and to estimate the intra-group light.
 Taking advantage of the deep and multiband photometry and of the large field of view of the
VST telescope, we studied the structure of the galaxy members and the
faint features into the intra-group space and we give an estimate of the intragroup diffuse light in the NGC 5018
group of galaxies. We found that $\sim$ 41\% of the total {\it g}-band
luminosity of the group is in the form of intragroup light (IGL).
The IGL has a {\it (g - r)} color consistent with those of
other galaxies in the group, indicating that the stripping leading to
the formation of IGL is ongoing. 
From the study of this group we can infer that there are at
least two different interactions involving the group members: one
between NGC 5018 and NGC 5022, which generates the tails and ring-like
structures detected in the light, and another between NGC 5022 and
MCG-03-34-013 that have produced the HI tail. A minor
merging event also happened in the formation history of NGC 5018 that have perturbed the inner structure of this galaxy.
\end{abstract}



\section{Introduction}

The hierarchical, merger-dominated picture of galaxy formation
predicts that the observed galaxies and their dark halo (DH) were
formed through repeated merging processes of small systems
\citep{DeLucia06}. Under this paradigm, clusters of galaxies are the
most recent objects to form and their central galaxies continue to
undergo active mass assembly and accretion of smaller groups.

Throughout this cluster assembly, individual galaxies interact
with one other. During these interactions, matter can be stripped from
galaxies and form tidal tails, shells, bridges, and liberating stars
from their host galaxies, which contribute to formation of a very faint
component of diffuse intacluster light (ICL) \citep{Mihos15}.

According to this picture, any star that becomes unbound from its host
galaxy contributes to the growth of ICL, which can then be considered
as the fossil record of all the past interactions. If so, the main properties
of the ICL, such as color, metallicity, spatial distribution, are
closely linked to the properties of galaxies in which intracluster
stars are originated, and then can help to disentangle their formation history.

In the group environment, interactions between galaxies are slow
leading to a strong tidal stripping of material and to the formation
of tidal debris, which are mixed into the ICL when the groups fall
into the cluster. In situ star formation can also occur in the
intracluster medium, due to the presence of gas stripped from
infalling galaxies, contributing to feed the ICL. Following this
evolutionary picture, the diffuse intragroup light (IGL) can be
considered as a precursor to the ICL in massive clusters of galaxies.

Since groups contain most of the galaxies in the Universe ($\sim$
60\%) at present day, the group environment is of particular interest
for the study of the intragroup starlight. IGL is in fact quite
evident in strongly interacting groups, given that the strong
interactions are able to expel diffuse material out to large distances
\citep{DaRocha05,Watkins15}.

In the last decade, a big effort was made to determine the amount and
spatial distribution of intragroup light in both normal and compact
groups, both on the observational and on the theoretical side.

Theoretical studies give conflicting predictions on the fraction of
the IGL component and of its variation with the group
mass. Numerical simulations by \citet{Napolitano03,Lin04,Murante04,Murante07,Purcell07,Watson12,Tollet17} predict a strong
evolution of the ICL and IGL fraction with the cluster and group mass,
while \citet{Krick07} found an anti-correlation between the ICL/IGL
fraction and the cluster/group mass. \citet{Contini14} found that the fractions of ICL and IGL
predicted by their models range between 10\% and 40\% and that they
don't vary as a function of the mass. Models by
\citet{Sommer06,Rudick06} show that from 12\% to 45\% of the light in groups is in the
form of IGL, and that the fractions of IGL/ICL can
be used as a ``dynamical clock'', since they
increase with the degree of dynamical evolution of the group/cluster
(i.e. more evolved groups/clusters have largest fractions of diffuse
light).

On the observational side, the diffuse IGL/ICL component has been mapped
  using several observational techniques,  such as  deep imaging  \citep{Feldmeier02, Mihos05, Zibetti05}, the detection of
  red giant branch stars associated with the diffuse stellar component
  \citep{Williams07}, and intracluster planetary nebulae
  \citep{Arnaboldi02, Arnaboldi04, Aguerri05, Aguerri06, Gerhard05,
    Gerhard07, Castro09, Longobardi13, Longobardi15}.

The fraction of IGL has been estimated in
some compact groups of galaxies. \citet{DaRocha05, DaRocha08} found
smooth envelopes of diffuse light around the studied galaxy groups,
and estimated IGL fractions of 11\%-46\%, with colors compatible with
those of galaxies in the groups. They also suggest an evolutionary
sequence for the analyzed groups, i.e. groups with highest IGL
fraction are in a more advanced phase of their dynamical evolution,
with respect to groups with smaller fractions of IGL. Also
\citet{White03} found a consistent amount of IGL (38\%-48\%) in the
compact group HCG 90, with a color distribution consistent with an old
stellar population, while \citet{Aguerri06} found that an IGL
fraction for the group HCG 44 of 4.7\%. In contrast, \citet{Watkins17} found no diffuse
light in the M96 group. \citet{DeMaio18} by studying the properties of 23 galaxy
groups found that the Bright Cluster Galaxies (BCGs) together with the
ICL constitute a higher fraction of the total mass in groups than in
clusters, and that the group environments are more efficient ICL
producers.
\citet{McGee10} by studying
spectroscopically confirmed intragroup supernovae Ia, inferred that 47\%
of the stellar mass in the analyzed galaxy groups was in the form of
IGL.

The study of intracluster planetary nebulae in two fields of the Virgo cluster,
  by \citet{Arnaboldi04} , revealed that the diffuse light contributes
  from 17\% to 43\% of the total (i.e.,
galaxies + diffuse light) luminosity density in the imaged fields.
\citet{Castro03}, by carrying out a wide field survey for
  emission line objects associated with the intragroup HI cloud in the
  Leo group, have been able to set an upper limit for the ratio of diffuse intragroup to galaxy
light that is $\sim$ 1.6\%. \citet{Castro09} by using planetary nebulae
detected in several regions of the Virgo cluster, derived a
fraction of ICL of $\sim$ 7\% of the total light in Virgo cluster
galaxies.

Recently, wide-field and deep photometric studies have revealed the
presence of diffuse IGL even in the less dense environment of the
loose groups \citep{Ibata14,Okamoto15}, where slow encounters are particularly effective at liberating stars from galaxies in the
intragroup medium.
Loose groups of galaxies are intermediate in scale between
galaxies and rich clusters, and due to their irregular nature their
definition is ambiguous. Their environment is also intermediate between that
of isolated galaxies and that of the cores of rich clusters, and
therefore the study of IGL in these environments is of particular
interest, given that it can give clues on galaxy and galaxy cluster
evolution.

Despite these many programs aimed at the study of the IGL/ICL in a
variety of different systems in different dynamical states, there are
still observational and interpretational difficulties to still face,
which make the investigation of the IGL and ICL not a trivial
task. Over the past decade many observational studies of these diffuse
components have been carried out, but despite this the amount of ICL
is hard to estimate. First,
the fact that ICL and IGL components are diffuse and have very faint
levels of surface brightness, which implies accurate techniques to
correctly estimate all source of contamination, first the
background. Second, the fact that the IGL/ICL is strictly connected to
the BCGs and it is not always possible to distinguish from photometry
only the diffuse (unbound) component from the extended (likely bound)
stellar halos of the BCGs. 


In the recent years, a great impulse to these researches has been
given by deep photometric surveys aimed at studying galaxy structures
out to the regions of the stellar halos, where the galaxy light merges into the intra-cluster component \citep{Ferrarese12,Duc15,Munoz15,Merrit16}.
The VST Early-type Galaxy  Survey (VEGAS)
(see http://www.na.astro.it/vegas/VEGAS/Welcome.html) is just a part of this campaign and it is producing competitive results.
From the surface photometry for the cD galaxy NGC~4472,  a tail of intracluster light was detected between $5R_e\le R \le 10R_e$, in the
range of surface brightness of $26.5 - 27.6$ mag/arcsec$^2$ in the $g$  band \citep{Cap15}.
New results on six massive early-type galaxies (including NGC~4472) in the VEGAS, confirm the feasibility of such a survey to reach the faint
surface brightness levels of $27 - 30$~mag/arcsec$^2$ in the $g$ band, out to $\sim10 R_e$  \citep{Spavone17}.
Therefore, taking advantage of the deep photometry, the build up history of the stellar halo can be addressed by comparing the surface
brightness profile and the stellar mass fraction with the prediction of cosmological galaxy formation.

As part of VEGAS, the {\it Fornax Deep Survey (FDS)} at VST  covers the Fornax cluster out to the virial radius ($\sim0.7$~Mpc), with an area
of about 26 square degrees around the central galaxy NGC~1399, and including the SW subgroup centered on Fornax~A.
First results have provided the up-to-date largest mosaics that covers an area of $3\times6$ square degrees around the central galaxy NGC~1399 (see the ESO photo release at https://www.eso.org/public/news/eso1612/) and an area of  $\sim 4 \times 2$~square degrees around the central galaxy NGC~1316 of the SW group of the Fornax cluster \citep{Iodice16, Iodice17a}.
The deep photometry, the high spatial resolution of OmegaCam and the
large covered area allow  to map {\it i)}  the surface brightness
around NGC~1399 and NGC~1316 out to an unprecedented distance of
$\sim200$~kpc  down to $\mu_g \simeq 29-31$ mag/arcsec$^2$
\citep{Iodice16, Iodice17a};  {\it ii)} to trace the spatial
distribution of candidate Globular Clusters (GCs) inside
$\sim0.5$~deg$^2$ of  the cluster core \citep{Dabrusco16,Cantiello18};
{\it iii)} to detect new and faint  ($\mu_g \simeq 28-30$
mag/arcsec$^2$) features in the intracluster region between NGC~1399
and NGC~1387 \citep{Iodice16} and in the outskirts of NGC~1316 \citep{Iodice17a}; {\it iv)} to detect an unknown region of intra-cluster light (ICL) in the core of the cluster, on the West side of NGC 1399 \citep{Iodice17}.

Thanks to the coverage of  different morphological types, masses and
environments in VEGAS, we have started a  study dedicated to the
synoptic analysis of GC systems in different host galaxies. As an
example, preliminary results from GCs in NGC3115 and NGC1399, support
the scenario where the red/metal-rich GCs component is bound to the
galaxy, while the blue/metal-poor GCs are gravitationally associated
with a cluster-wide component, if present \citep{Cantiello18}.

In this work we present a new deep mosaic of $1.2 \times 1.0$ square degrees of the group of galaxies centered on NGC~5018, acquired at the ESO VST.
We use  {\it u, g, r} images to analyse the structure of the group members and to estimate the intra-group light.
This work represents a pilot study in the framework of the VEGAS survey to study the IGL
components in the less dense environments of the groups.


We adopt a distance for NGC 5018 of D = 40.9 Mpc \citep{Tully88},
therefore the image scale is 198.3 parsecs/arcsec.

The plan of the paper is the following. In Sec. \ref{lit} we present
the  properties  of NGC 5018 as found in the literature, while in
Sec. \ref{obs} we describe the data used in this work, as well as the
adopted observing strategy and the data reduction. In Sec. \ref{phot},
\ref{ir} and \ref{UV} the photometric optical, infrared and UV analysis are
described. In Sec. \ref{fit_sec} we present the 1D fitting procedure
adopted and the derived accreted mass fractions of galaxies in the
group. In Sec. \ref{conc} we draw our conclusions.

\section{NGC 5018 in the literature}\label{lit}
\begin{figure*}
\centering
\hspace{-0.cm}
 \includegraphics[width=13cm, angle=0]{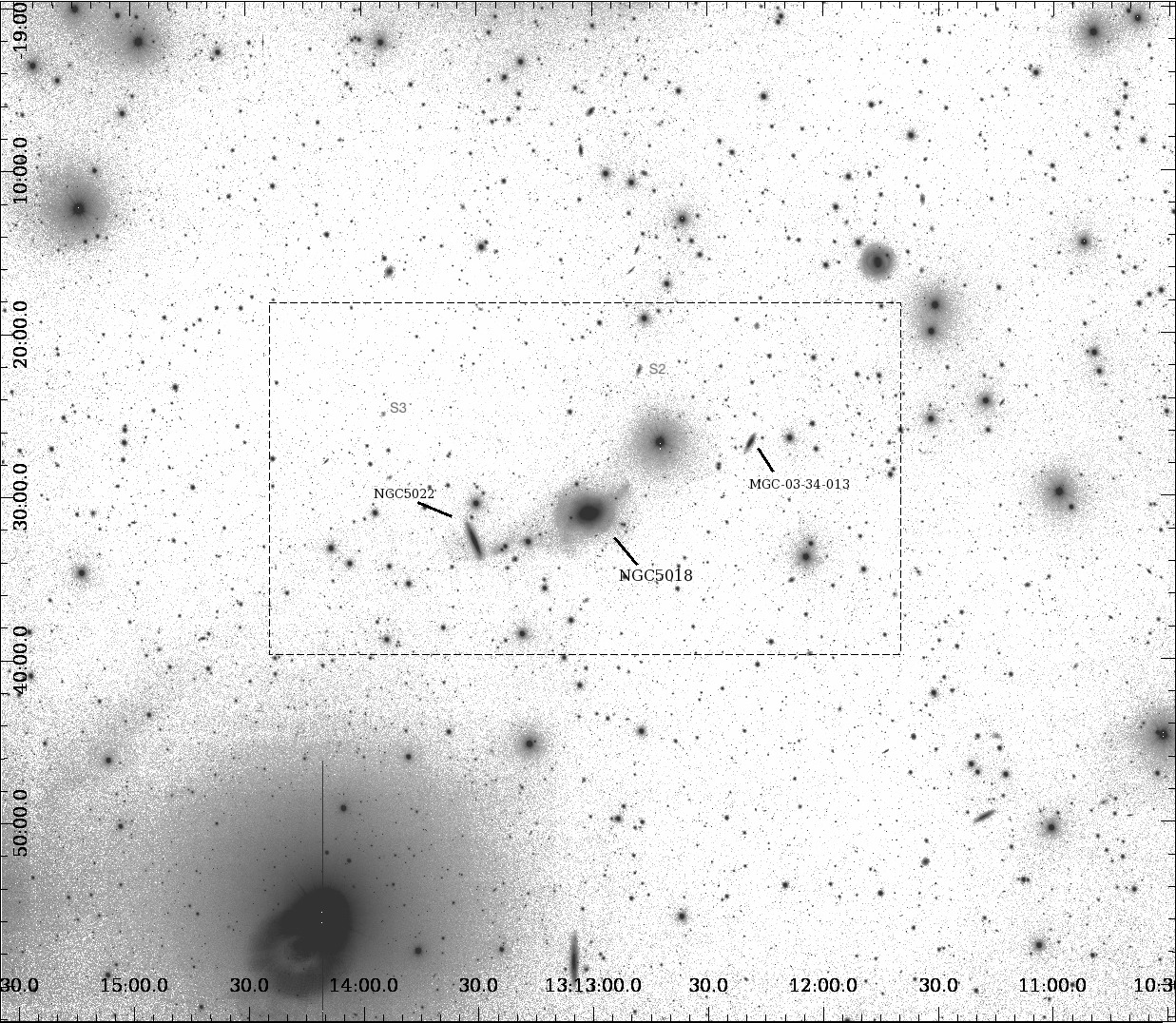}
 \includegraphics[width=13cm, angle=0]{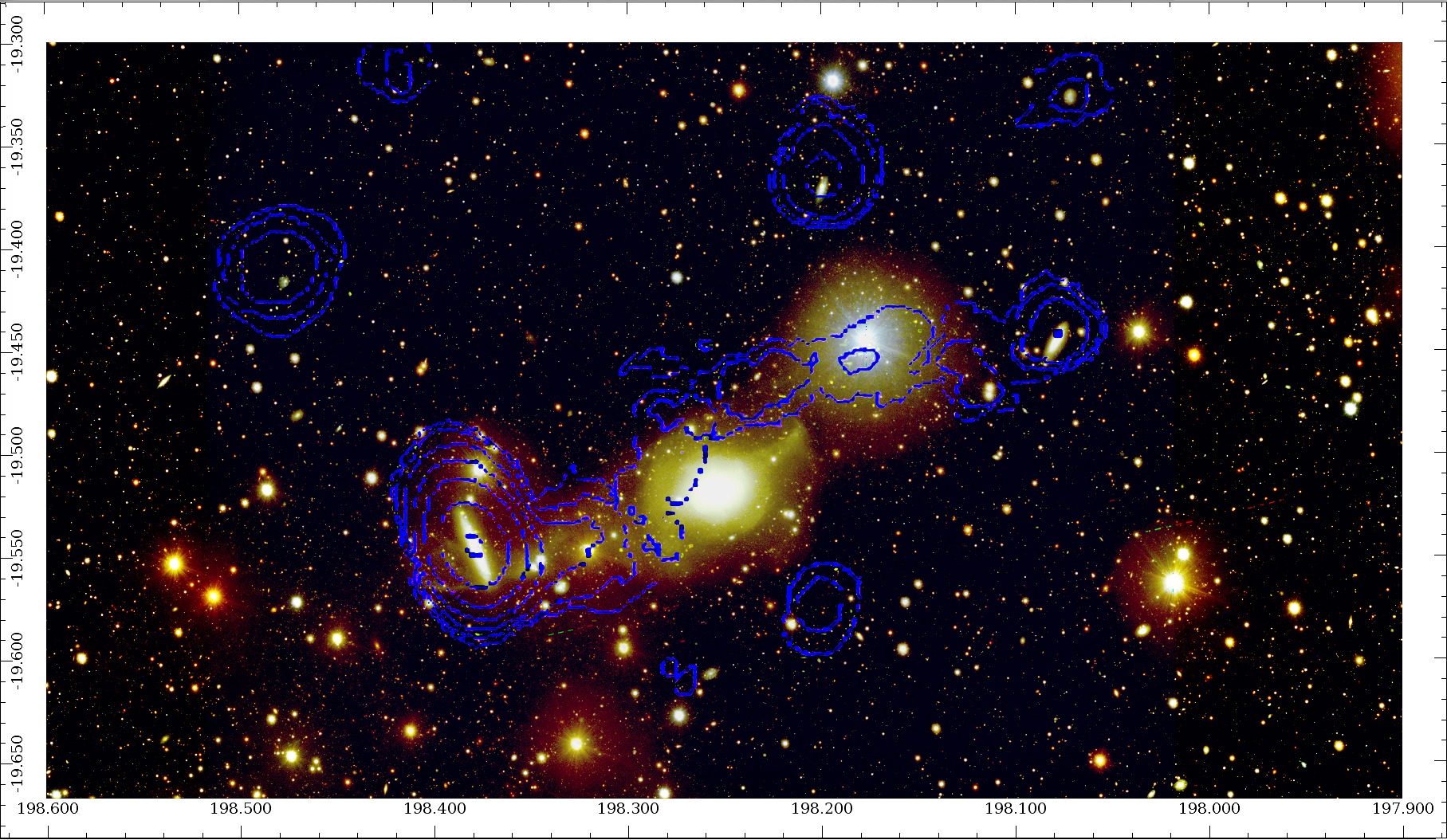}
\caption{{\it Top}: VST g-band mosaic of NGC 5018 group. The image
  size is 1.2$^{\circ} \times$ 1.0$^{\circ}$. {\it Bottom}: Color
  composite image of the central regions of the group (0.7$^{\circ}
  \times$ 0.4$^{\circ}$), assembled from {\it u}, {\it g} and {\it r}
  band VST images, with the HI
  map from the VLA superimposed (blue contours), adapted from \citet{Kim88}.}\label{field}
\end{figure*}

NGC 5018 is the brightest member of a small group of five galaxies \citep{Gourg92}, composed by the
edge-on spiral, NGC 5022, the S0 galaxy MCG-03-34-013, and the two face-on
dwarf, gas-rich ($\sim 10^{8} M_{\odot}$) spirals labeled as S2 and S3 by \citealt{Kim88} (see
Fig. \ref{field}). \citet{Kim88}, adopting a distance $D \sim 22.5
h^{-1}$ Mpc, also estimate that the
group has a size of $\sim$ 200 kpc and a total mass of $\sim 4 \times
10^{12} M_{\odot}$. Despite its classification as ``normal''
elliptical, NGC 5018 shows several signs of a past interaction event,
such as a very complex system of dust lanes, shells, a tail on the NW
side, and a prominent bridge of gas toward the companion galaxy NGC
5022. The galaxy is also present in the list of shell galaxies revealed by \citet{Malin83}.

The HI mass associated to NGC 5018, estimated by \citet{Kim88}, is
$\sim 4 \times 10^{8} M_{\odot}$, while the same authors found that
the HI in NGC 5022 is distributed in a rotating disk, with a mass of
$\sim 2 \times 10^{9} M_{\odot}$. 

Follow-up HI observations made by \citet{Guhathakurta90}, showed that
the HI bridge actually connects NGC 5022 with MCG-03-34-013, while it
bifurcates at the position and radial velocity of NGC 5018. They also
estimate that the interaction which led to the formation of the
northern plume was recent ($\sim 6 \times 10^{8}$
Gyrs). \citet{Guhathakurta90} assert that their HI and optical data
provide the direct observational evidence for the formation of a shell
system, through the merger of an elliptical galaxy with a cold disk system.

Many authors also found the presence of a young ($\sim$ 3 Gyrs) and dominant
stellar population in the central regions of NGC 5018
\citep{Bertola93,Carollo94,Leonardi00,Buson04,Rampazzo07}, with a
near solar metallicity. The peculiarities found in NGC 5018, such as
the lower metallicity with respect to the giant ellipticals, the
presence of an $Mg_{2}$ index much weaker than those of ellipticals of
similar absolute magnitude, the low UV flux level, the presence of
bluer shells, a complex system of dust lanes, and a bridge of HI
connecting NGC 5018 with NGC 5022, pose serious problems to the
classical picture of elliptical galaxies formation. 

\citet{Bertola93} proposed two possible scenarios for the formation of
this peculiar galaxy. The first scenario is the merging of many
separate smaller galaxies, having low metallicities, while the other
possibility is the formation ab initio with low metallicity and the
subsequent, recent, merging with a small disk galaxy, which is able
to reproduce the morphological peculiarities of NGC
5018. \citet{Bertola93} discounted the dust obscuration as the cause
of the low UV flux level because the best photometry available at that
time showed that the dust lane does not cover the regions of their UV
spectra. Later, \citet{Carollo94} showed the presence of the dust till
to the very central regions, demonstrating that the reddening effect
was not negligible.

\citet{Leonardi00} derived a robust reddening insensitive estimate of the age
of the young stellar population in the central regions of NGC
5018. The estimated age of 2.8 Gyrs of this population, which
dominates the visible part of the spectrum, also explain the small UV
flux with no need to invoke a large amount of dust obscuration.

NGC 5018 has been examined by \citet{Rampazzo13}
using Spitzer-IRS spectrograph in Infrared, in the range 4-38 $\mu$m
(see in particular their Figure 5).
The galaxy nucleus shows Polycyclic Aromatic Hydrocarbon features (PAHs)
with interband ratios typical of late-type galaxies. These PAH
features are present only in 9$^{+4}_{-3}$ fraction of early-type galaxies 
in the Revised Shapley-Ames catalogue examined by \citet{Rampazzo13}.
This  suggests that NGC 5018 has a still actively star forming nuclear region,
since IRS (SL+LL) spectra integrate in a projected region of  0.7$\times$3.3 kpc.

\citet{Hilker96} studied the properties of the globular clusters (GCs)
of NGC 5018. They detect a poor globular cluster system, that can be
divided in two populations: a small population of blue GCs, with ages
between several hundred Myrs and 6 Gyrs, and one or two populations of
older GCs. The first population was probably formed during the last
interaction, while the second one is associated to the original
galaxy. Another peculiarity of this galaxy is the distribution of GCs;
\citet{Hilker96} found that blue GCs are missing in a stripe
extending from South-East to North of the galaxy and that red GCs are
instead overabundant in this same stripe. The inhomogeneous
distribution of the HI found by \citet{Guhathakurta90}, could be the
responsible of the inhomogeneous distribution of GCs, but this can not
explain the overabundance of red objects in the stripe. The small amount of
young GCs found in NGC 5018 ($\sim$ 10 \%) is an evidence against a
significant increase of GCs by a merger event for this galaxy.

NGC 5018 was also observed in the X-rays with the {\it Chandra X-Ray
  Observatory}'s Advanced CCD Imaging Spectrometer by
\citet{Ghosh05}. These authors found six nonnuclear X-ray point
sources in NGC 5018, as well as diffuse hot gas that may be the
remnant of interactions of NGC 5018 with its neighbor
galaxies. Their  total  absorption-corrected luminosity  for  the
diffuse  light  was  13.7 $\pm$ 1.5 $\times$ 10$^{39}$ erg/s.
\citet{Ghosh05} also estimated a radiative cooling time for
the hot plasma of a few times 10$^{7}$-10$^{8}$ yr, which turns to be
shorter than the age of the last interaction
\citep{Guhathakurta90}. They conclude that the gas is still falling in
and that it is reheated by ionization, stellar winds, and supernovae
by recent star formation activity. Moreover, they found that, even if
there is little current star formation in NGC 5018, there is a
significant reservoir of gas which maintains a low but steady level of
star formation, also explaining the diffuse X-ray emission of the
galaxy.

Recently, {\it Chandra} X-ray data have been analyzed by
\citet{Smith18} to study the diffuse X-ray emitting gas in major
mergers. They did a rough classification of the systems in their
sample into seven merger stages. According to the adopted criteria,
they classify NGC 5018 as a very late stage major merger remnant.

Summarizing, our present knowledge converges in indicating that in NGC 5018 merging episode/s may have characterized its recent history. With the study of the IGL we aim at adding a new tessera in the mosaic considering its evolutionary environment. 

\section{Observations and data reduction} \label{obs}

The data presented in this work are part of the VEGAS\footnote{http://www.na.astro.it/vegas/VEGAS/Welcome.html} survey,
which is a multiband {\it u, g, r} and {\it i} imaging
survey, obtained with the ESO VLT Survey Telescope (VST). 

With VEGAS we are mapping the light distribution and colors out to
8-10 Re and down to $\mu \sim$ 30~mag/arcsec$^{2}$  in the $g$ band, for a large sample ($\sim 42$) of early-type galaxies in different environments, including giant cD galaxies in the core of clusters.
The main science goal of VEGAS are to study 
{\it i)} the galaxy structure and its faint stellar halo, including the diffuse light component, inner substructures as signatures of recent cannibalism events, inner disks and bars; 
{\it ii)}  the external low-surface brightness structures of the galaxies (tidal tails, stellar streams and shells) and the connection with the environment;
{\it iii)} for those galaxies in the sample with $D < 40$~Mpc, the GCs and galaxy satellites in the outermost regions of the host galaxy and their photometric properties (e.g. GC colours and mean GC radial colour changes).
A more detailed description of the survey, the selected targets and the main scientific aims can be found in \citet{Cap15}.
The data reduction, performed by using the VST-Tube pipeline
\citep{Grado12}, and the
analysis are described in details by \citet{Cap15,Spavone17,Iodice16,Iodice17}.

The most important step of the data processing is the
estimation and subtraction of the sky background.
For this reason, we adopt a {\it step dither} observing strategy for galaxies
with large angular extent, since it allows a very accurate
estimate of the sky background
\citep{Spavone17,Iodice16,Iodice17}. This strategy consists of a cycle of short exposures centered
on the target and on offset fields ($\Delta \sim \pm 1$ degree). With such
a technique the background can be estimated from exposures taken as
close as possible, in space and time, to the scientific images. This ensures
better accuracy, reducing the uncertainties at very faint surface brightness
levels. This observing strategy
allowed us to build an average sky background of the night, which was subtracted
from each science image. This average sky frame takes into account the small contribution to the sky
brightness by the smooth components plus  the extragalactic background light.
The residual fluctuations in the background are then estimated, and taken into account,
by using the methodology described by \citet{Pohlen06}, as explained
in \citet{Spavone17}. These
fluctuations of the sky background have been taken into account in the error
estimates we quote on our surface brightness measurements.

In the case of NGC 5018 we adopted as offset field the one located at
the West side of the galaxy (R.A. 13h08m47.260s Dec. -19d20m46.40s), since this field is not very crowded and,
mainly, because it does not contain very bright galaxies and stars.

The data used in this paper consist of exposures in {\it u}, {\it g} and {\it
  r} bands obtained with VST + OmegaCAM, in
visitor mode (run ID: 096.B-0582(B), 097.B-0806(A) and
099.B-0560(A)), during dark time, in photometric conditions, with an
average seeing between 0.6 and 1.1 arcsec (see Tab. \ref{data}). In this work we analyze an
area of 1.2 square degrees around NGC 5018. This area covers the three
main galaxies of the small group, composed by the dominant early type
galaxy, NGC 5018, the
edge-on spiral, NGC 5022, the S0 galaxy MCG-03-34-013, analyzed in
this work, whose main properties are listed in Tab. \ref{basic}.

\begin{table}
\caption{\label{data}VST exposures used in this work.} \centering
\begin{tabular}{lccccc}
\hline\hline
Object & Band & $T_{exp}$ & FWHM\tablenotemark{a}\\ 
    & & [sec] & [arcsec] & \\
\hline \vspace{-7pt}\\
NGC 5018 & u & 8280 & 0.77\\ 
                & g & 8280 & 0.77\\ 
               & r & 8280 & 0.94\\ 
\hline
\end{tabular}
\tablenotetext{a}{Median value of the FWHM.}
\end{table}

\begin{table}
\setlength{\tabcolsep}{2.1pt}
\caption{\label{basic}Basic properties of the galaxies studied in this paper.} \centering
\begin{tabular}{lccc}
\hline\hline
Parameter & Value & Ref.\\
\hline \vspace{-7pt}\\
 & {\it NGC 5018} & \\
\hline \vspace{-7pt}\\
Morphological type & E3 & RC3 \\
R.A. (J2000)& 13h13m01.0s& NED \\
Dec. (J2000) &  -19d31m05s & NED\\
Helio. radial velocity & 2816 km/s&NED\\
Distance& 40.9 Mpc& \citet{Tully88} \\
\hline \vspace{-7pt}\\
 & {\it NGC 5022} & \\
\hline \vspace{-7pt}\\
Morphological type & SBb pec & RC3 \\
R.A. (J2000)& 13h13m30.8s & NED \\
Dec. (J2000) & -19d32m48s & NED\\
Helio. radial velocity & 3001 km/s&NED\\
Distance& 40.4 Mpc& NED \\
\hline \vspace{-7pt}\\
 & {\it MCG-03-34-013} & \\
\hline \vspace{-7pt}\\
Morphological type & S0 & RC3 \\
R.A. (J2000)& 13h12m18.9s & NED \\
Dec. (J2000) & -19d26m46s & NED\\
Helio. radial velocity & 2691 km/s&NED\\
Distance& 40.7 Mpc& NED \\
\hline \vspace{-7pt}\\
\hline
\end{tabular}
\end{table}


\section{Surface photometry}\label{phot}

\subsection{Isophotal analysis}\label{iso}

The isophotal analysis of the galaxies in this work is performed on the final mosaic in each band with the {\small IRAF}\footnote{IRAF ({\it
Image Reduction and Analysis Facility}) is distributed by the National
    Optical Astronomy Observatories, which is operated by the
    Associated Universities for Research in Astronomy, Inc. under
    cooperative agreement with the National Science Foundation.} task
  {\small ELLIPSE}.

\begin{figure*}
\centering
\hspace{-0.cm}
 \includegraphics[width=13cm, angle=0]{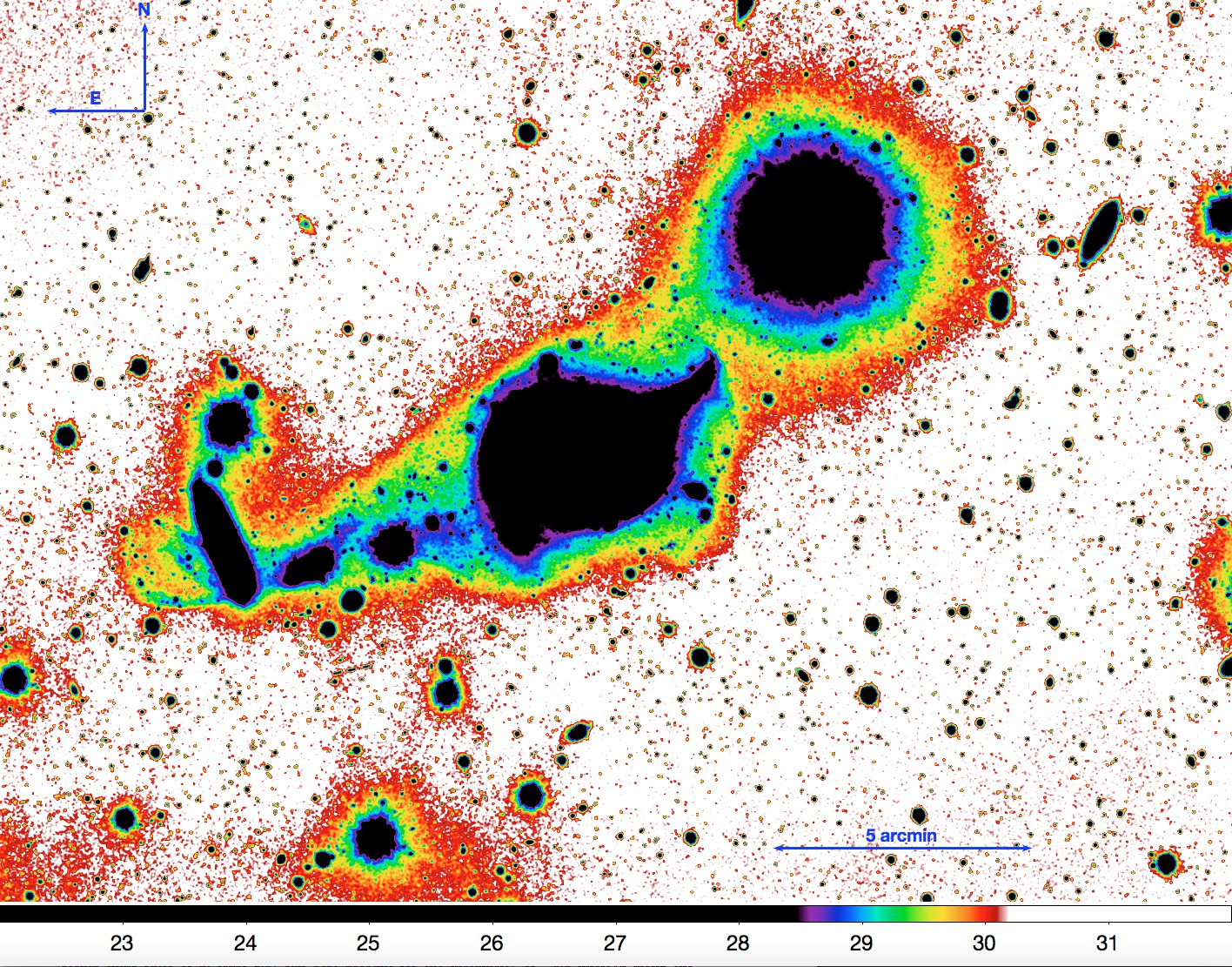}
\plottwo {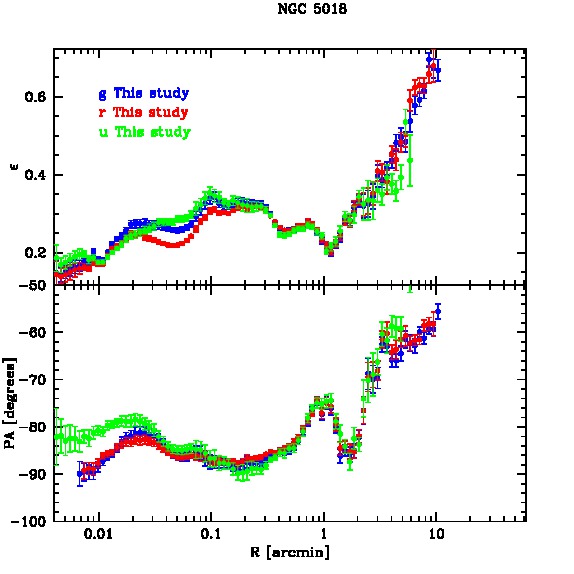}{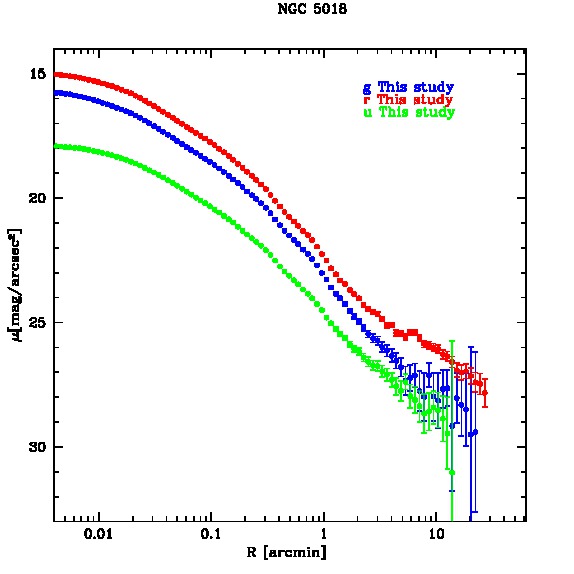}

\caption{{\it Top}: VST {\it g} band sky-subtracted image of the $0.4^{\circ}
\times 0.4^{\circ}$ field around NGC 5018. The color scale represents
surface brightness in mag/arcsec$^{2}$. {\it Bottom left}: ellipticity
($\epsilon$) and position angle (P.A.) profiles for NGC 5018, in the {\it g}
band (blue points), {\it r} band (red points), and {\it u} band (green
points). {\it Bottom right}: Azimuthally averaged
surface brightness profiles in the {\it g} (blue), {\it r} (red), and
{\it u} (green)
bands.}\label{N5018}

\end{figure*}

In the top panel of Fig. \ref{N5018} we show the VST {\it g} band
image of NGC 5018, NGC 5022 and MCG-03-34-013, in
surface brightness levels. In the bottom-left panel of the same figure we show the ellipticity
($\epsilon$) and position angle (P.A.) profiles in the
{\it g} band (blue points), {\it r} band (red points), and {\it u}
band (green points) resulting from our isophotal
analysis, performed via the {\small ELLIPSE} task as described in
details in \citet{Spavone17}. The bottom-right panel shows the  {\it
  g}, {\it r} and {\it u} band VST azimuthally averaged surface
brightness profiles of NGC 5018. No correction for seeing blurring is applied to the inner
regions of the profiles.

\begin{figure*}
\centering
\hspace{-0.cm}
\plottwo {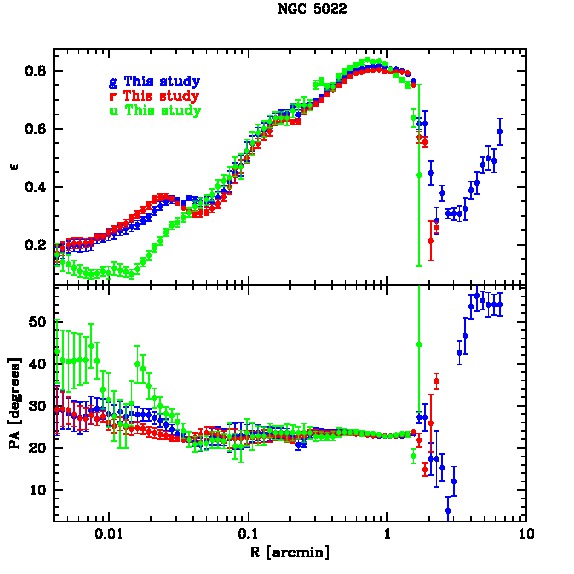}{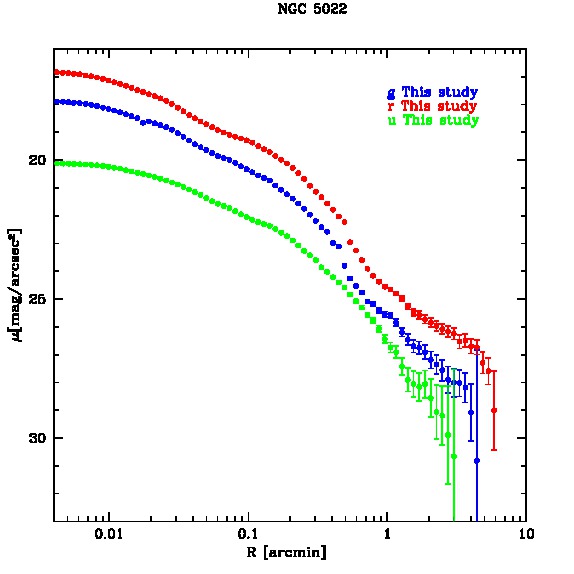}

\caption{{\it Left}: ellipticity
($\epsilon$) and position angle (P.A.) profiles for NGC 5022, in the {\it g}
band (blue points), {\it r} band (red points), and {\it u} band (green
points). {\it Right}: Azimuthally averaged
surface brightness profiles in the {\it g} (blue), {\it r} (red), and
{\it u} (green)
bands.}\label{N5022}

\end{figure*}

\begin{figure*}
\centering
\hspace{-0.cm}
\plottwo {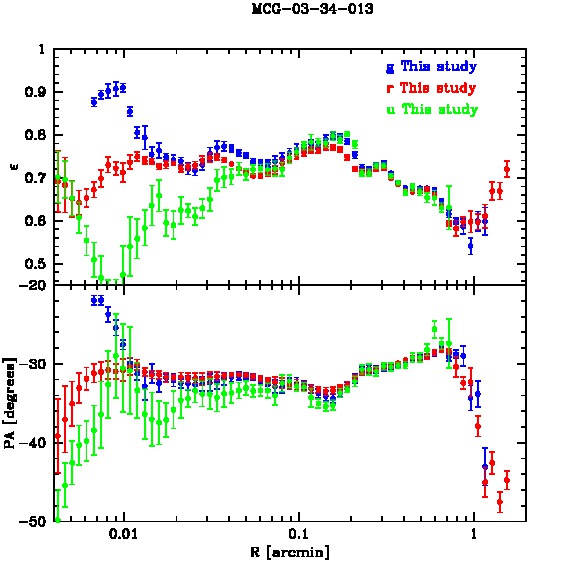}{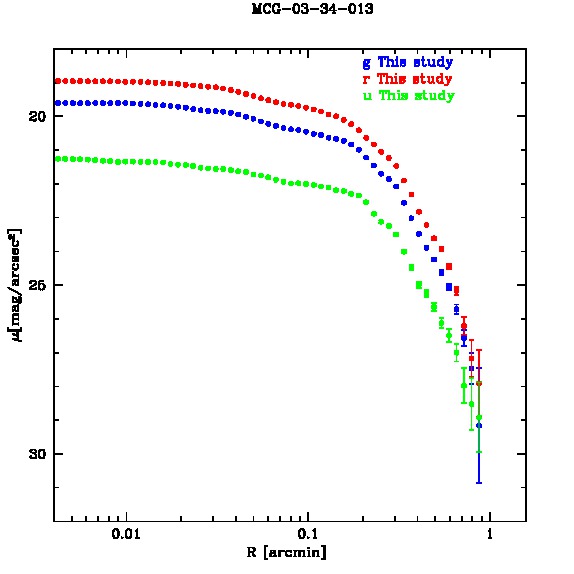}

\caption{{\it Left}: ellipticity
($\epsilon$) and position angle (P.A.) profiles for MCG-03-34-013, in the {\it g}
band (blue points), {\it r} band (red points), and {\it u} band (green
points). {\it Right}: Azimuthally averaged
surface brightness profiles in the {\it g} (blue), {\it r} (red), and
{\it u} (green)
bands.}\label{mcg}

\end{figure*}

The same isophotal analysis has been performed also for NGC 5022 and
MCG-03-34-013, and the results are shown in Fig. \ref{N5022} and
\ref{mcg}, respectively.

The growth curves obtained by isophote fitting, have been
extrapolated to the derive total magnitudes $m_{T}$, effective radii
$R_{e}$ and corresponding effective magnitude $\mu_{e}$ in each band (Tab.
\ref{par}). All the reported magnitudes are corrected for interstellar
extinction, by using extinction coefficients derived by \citet{Burstein82}.

\begin{table*}
\setlength{\tabcolsep}{2.1pt}
\scriptsize
\caption{Distances and photometric parameters for the
  sample galaxies.} \label{par}
\vspace{10pt}
\begin{tabular}{lcccccccccccccccc}
\hline\hline
Object & $D$ &$A_{u}$&$A_{g}$&$A_{r}$&$\mu_{e,u}$ & $r_{e,u}$&$m_{T,u}$& $M_{T,u}$&$\mu_{e,g}$ &
$r_{e,g}$&$m_{T,g}$& $M_{T,g}$& $\mu_{e,r}$ &
$r_{e,r}$&$m_{T,r}$& $M_{T,r}$\\
    &[Mpc] &[mag]&[mag] &[mag] & [mag/arcsec$^{2}$] & [arcmin] &[mag] &[mag] & [mag/arcsec$^{2}$] & [arcmin] &[mag] &[mag]& [mag/arcsec$^{2}$] & [arcmin] &[mag] &[mag]\\
&(a)&(b)&(b)& (b)&(c)&&(c)&(c)&(c) &&(c)&(c) &(c) &&(c)&(c)\\
\hline \vspace{-7pt}\\
NGC 5018    &    40.9    &   0.493   & 0.362& 0.263&  24.98    &  1.40   &    11.71& -21.35  & 21.92  &    0.54   &10.93&  -22.49  &24.68  & 3.41&9.29 &-23.77\\
 NGC 5022    &    40.4   &   0.513  &   0.377& 0.274&   23.83    &  0.43   &    14.09& -18.94  & 21.67  &    0.29   &13.38&  -19.65  &25.47  & 1.89&11.93 &-21.10\\
 MCG-03-34-013 &40.7&  0.495  &   0.364& 0.264&   22.33    &  0.23   &    15.50& -17.55  & 21.05  &    0.22   &14.19&  -18.86  &20.45  & 0.22&13.61 &-19.44\\
\hline
\end{tabular}
\tablenotetext{a}{Distance of NGC 5018 is
  from \citet{Tully88}; distances of NGC 5022 and MCG-03-34-013 are from
  NED.}
\tablenotetext{b}{Extinction correction in the {\it
    u}, {\it g}
  and {\it r} band are from \citet{Burstein82}.}
\tablenotetext{c}{Derived by integrating the growth curves and corrected for interstellar extinction.}
\end{table*}

\subsection{Two dimensional model of the light
  distribution}\label{model}

\begin{figure*}
\centering
\hspace{-0.cm}
 \includegraphics[width=15cm, angle=0]{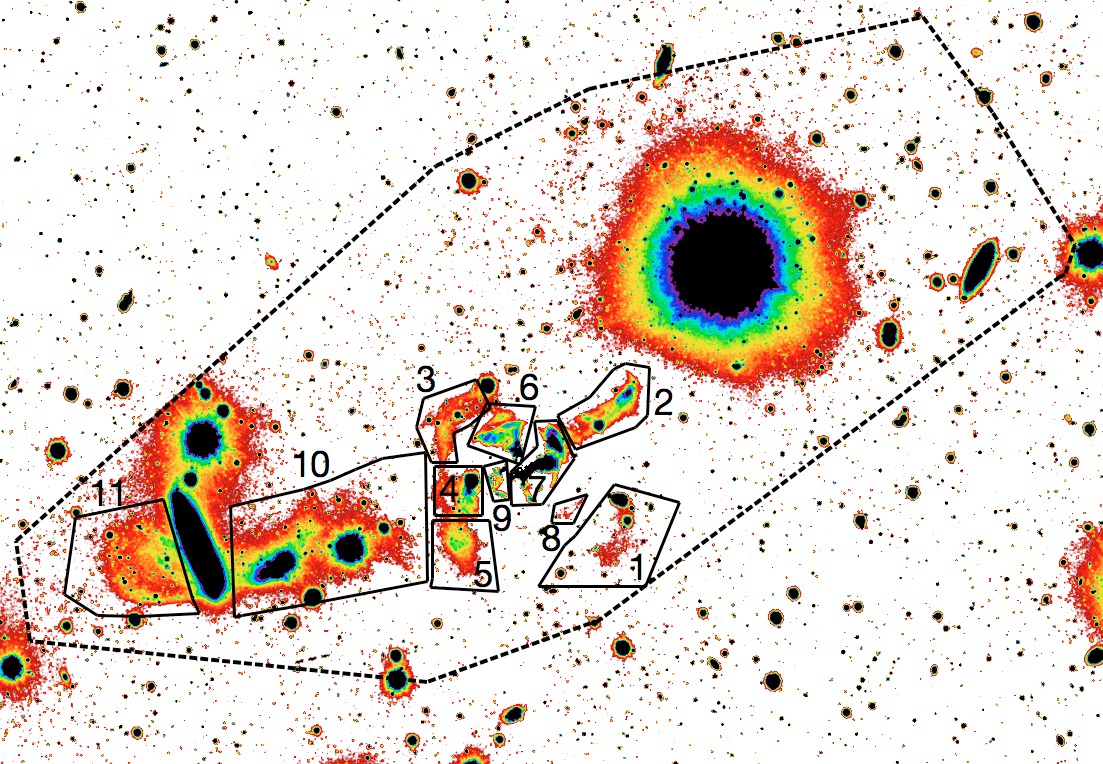}
\caption{Residual image obtained by subtracting from the VST {\it g}
  band image the galaxy model.
The numbers indicate the most luminous substructures for which we have
estimated the integrated colors (see Sec. \ref{color}).
The dashed polygon marks the area over which we estimated the
integrated magnitudes and colors of the intra group component (see
Sec. \ref{icl}.}\label{res}
\end{figure*}

The IRAF task BMODEL creates a 2-dimensional image
file containing a noiseless photometric model of a source image
(``parent image''). The
model is built from the results of isophotal analysis generated by the
isophote fitting task, ELLIPSE.
We use BMODEL to create a model of NGC 5018, which is able to
take into account also ellipticity and P.A. variations. In
Fig. \ref{res} we show the residual image obtained by subtracting from the VST {\it g}
  band image the galaxy model.

The complex structure of NGC 5018 stands out very clear from the
residual image. From this map, in fact, we can clearly identify a
numer of substructures, such as a prominent tail in the NW side of the
galaxy, the bridge connecting NGC 5018 with NGC 5022 and the loop on
the SE side of NGC 5022, which were also visible in the ``parent
image'', as well as multiple shells, filaments and fans of stellar
light, extending to the central galaxy regions.

\begin{figure}
\centering
\hspace{-0.cm}
 \includegraphics[width=9cm, angle=0]{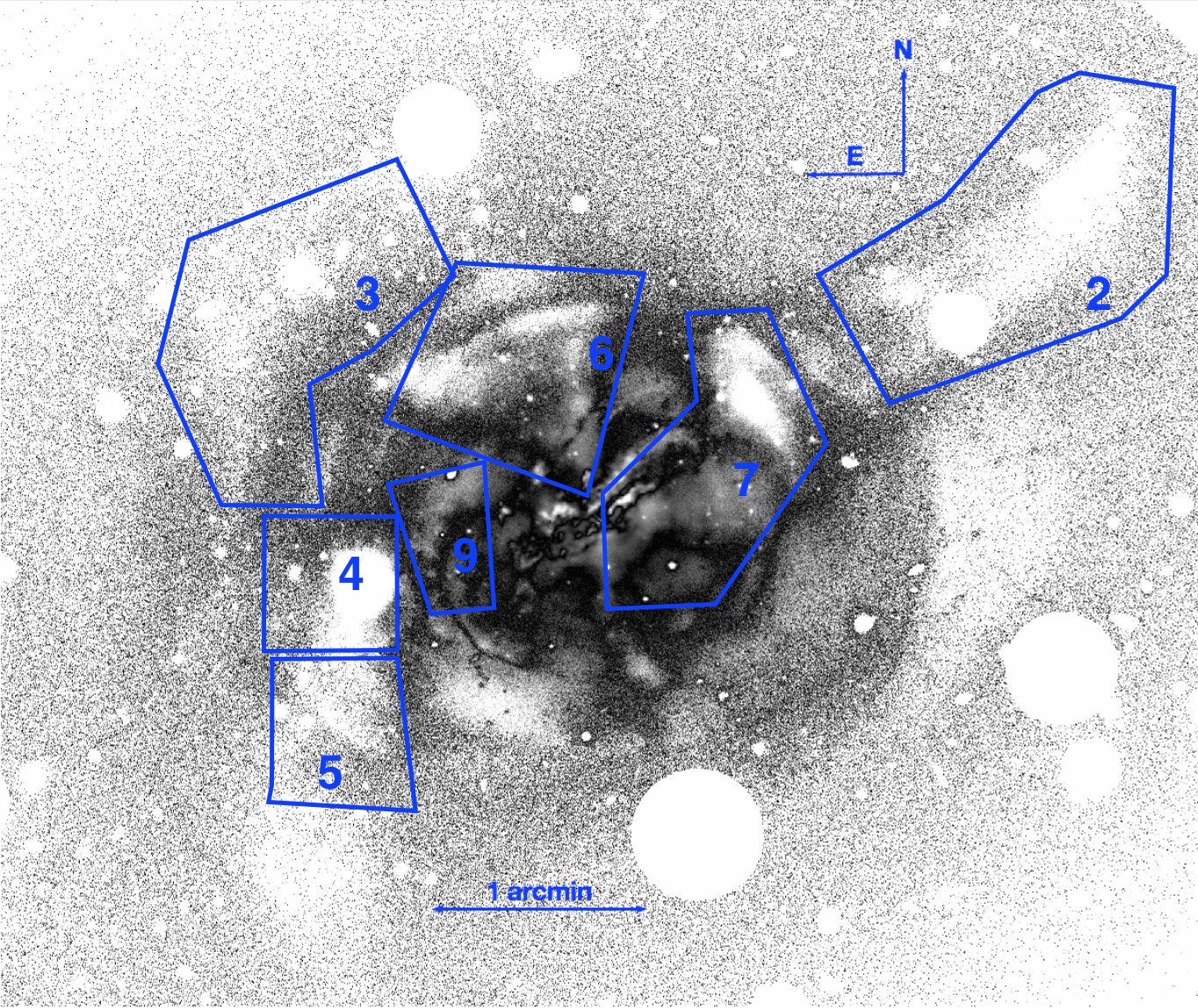}
\caption{Result of model fitting and division method employed to
  quantify the tidal disturbance of NGC 5018 (see text). The image size is 6
  $\times$ 5 arcmin. The overplotted blue contours mark the substructures
  shown in Fig. \ref{res} which are also visible in this image.}\label{tidal}
\end{figure}

We used the method described by \citet{Tal09} to quantify the tidal
disturbance of NGC 5018, by dividing the masked VST {\it g} band image
of the galaxy and the galaxy model obtained as explained above. The
resulting model frame is shown in Fig. \ref{tidal}

From this figure it appears even more clearly the complex dust lanes
system in the central regions of the galaxy, as well as its multiple
shells and fans of diffuse material. NGC 5018, in fact, has been
classified by \citet{Tal09} as an ``Highly Disturbed Galaxy'', on the
basis of its tidal parameter. According to these authors, such a
complex  system is  probably  the result  of  mergers  with
multi-component objects such as S0 or spiral galaxies.

\subsection{Color distribution}\label{color}

\begin{figure*}
\hspace{-0.cm}
 \includegraphics[width=19cm, angle=0]{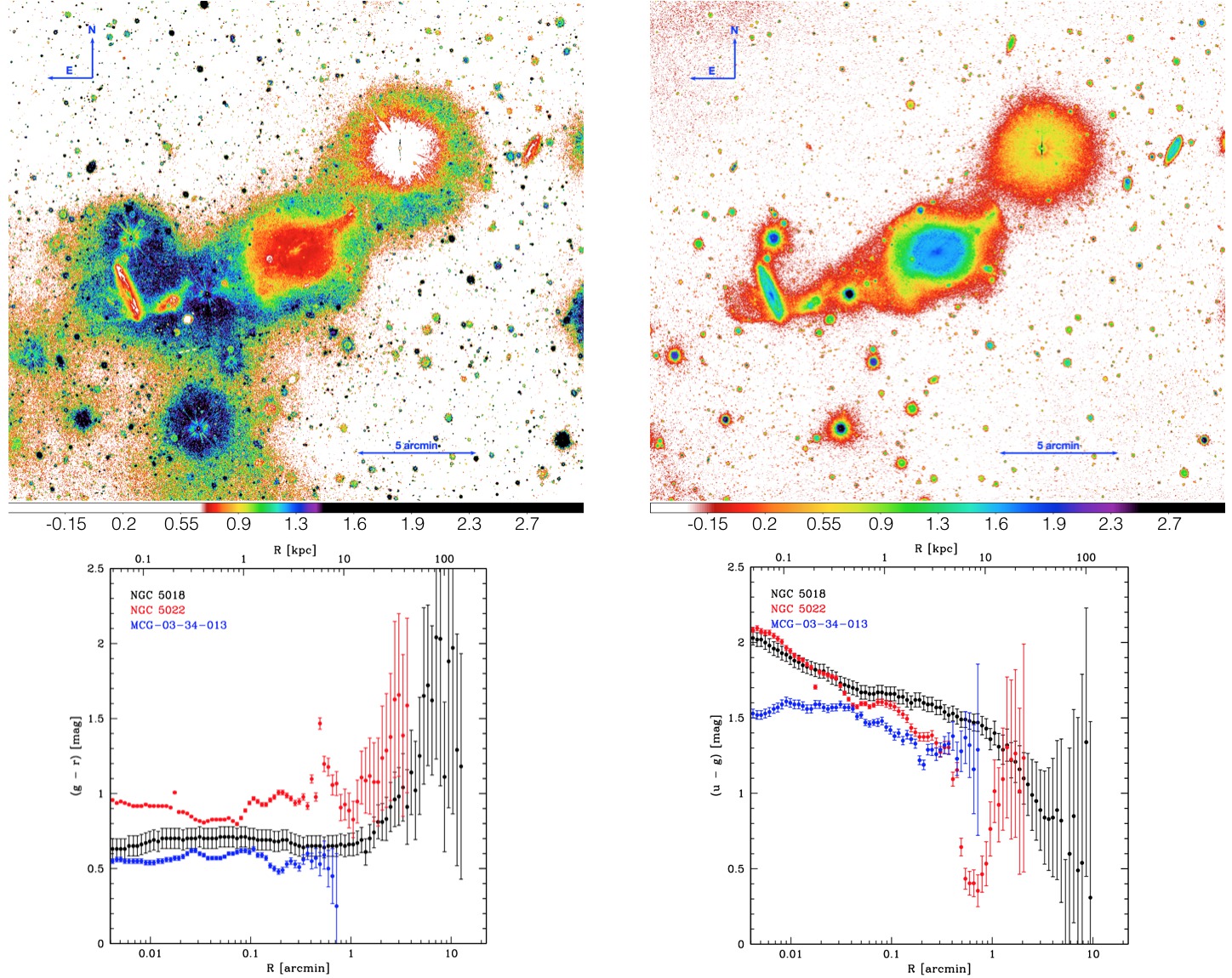}
\caption{{\it (g - r)} ({\it Top-left panel}) and {\it (u - g)} ({\it
    Top-right panel}) color maps centered on NGC 5018. {\it
    Bottom}: Azimuthally averaged, extinction corrected, {\it (g - r)} (left) and {\it (u - g)} (right)
  color profiles of NGC 5018 (black), NGC 5022 (red) and MCG-03-34-013
  (blue).}\label{col_5018}
\end{figure*}

We have derived the azimuthally averaged, extinction corrected {\it (g
  - r)} and {\it (u - g)} color profiles (Fig. \ref{col_5018}, bottom panel) and
the 2D color maps centered on NGC 5018 (top and middle panels). On average, central regions are bluer
then the outer ones, but the colors are consistent with those
typically found for ETGs \citep{LaBarbera12}. From both the color maps it is evident the complex system
of dust lanes in the center of the galaxy.

Azimuthally averaged, extinction corrected, {\it (g - r)} and {\it (u - g)} color profiles
have been derived also for NGC 5022 and MCG-03-34-013, and are shown
in the bottom panels of Fig. \ref{col_5018}.

\begin{figure}
\centering
\hspace{-0.cm}
 \includegraphics[width=9cm, angle=0]{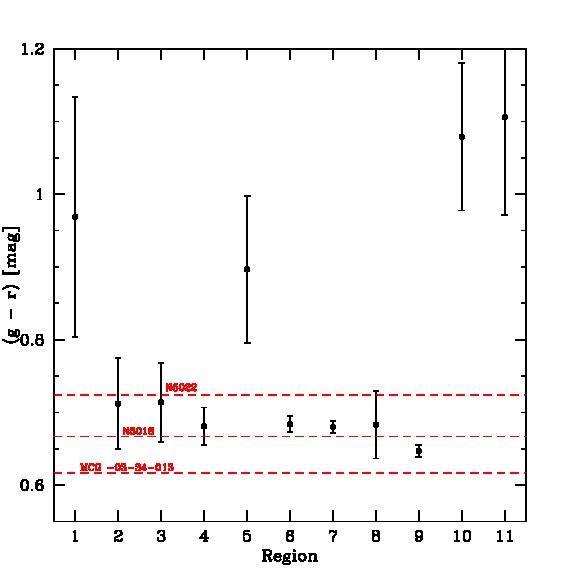}
\caption{
Integrated {\it (g - r)} colors in the 11
  regions marked on the residual image, Fig. \ref{res}. Red dashed lines represent the
  integrated colors of the three main galaxies in the VST field (see
  Fig. \ref{field}): NGC 5018, NGC 5022 and MCG-03-34-013.}\label{int_col}
\end{figure}

By using the residual map shown in Fig. \ref{res}, we derived the
integrated {\it  (g - r)} colors in the substructures standing out
from this map. To this aim, after masking all the bright sources, we define on the residual map a set of
polygons covering such substructures, and used them to estimate the
integrated colors, with the main goal to compare them with those of
the galaxies belonging to the group. In Fig. \ref{int_col} we show
the integrated colors of the different polygons, compared with those of
the galaxies ({\it (g-r)}$_{NGC 5018}$ = 0.67 $\pm$ 0.02, {\it (g-r)}$_{NGC
  5022}$ = 0.73 $\pm$ 0.02,
{\it (g-r)}$_{MCG-03-34-013}$ = 0.62 $\pm$ 0.02), marked as dashed red lines in the plot.


From this comparison, shown in Fig. \ref{int_col}, we can see that substructures labelled as
1 ({\it (g-r)} = 0.97), 5 ({\it (g-r)} = 0.9), 10 ({\it (g-r)} = 1.08), and
11 ({\it (g-r)} = 1.1) have colors consistent with each other, but
redder than those of the galaxies in the group. This suggests that all
the above features are part of a single structure, which
could have an external origin, since its color is not compatible with any of the galaxy in the
group. Moreover, we can note that the colors of these
substructures are comparable with those of the outskirts of both NGC
5018 and NGC
5022. The implications of this result will be discussed in Sec. \ref{conc}.

The colors of all the other identified substructures appear to
be similar to that of NGC 5018, suggesting that they could be part of
the parent galaxy.

\subsection{Integrated colors of the intra group light}\label{icl}

We used the IRAF task POLYMARK to define a region in
the residual map centered on NGC 5018 (see Fig. \ref{res}) covering the bulk of intra group light
(IGL). Then we use POLYPHOT to derive the integrated
magnitudes in the {\it g} and {\it r} bands, and {\it (g - r)} color inside this area. Foreground and
background bright sources have been masked and excluded from the
estimate of the integrated quantities (see also \citealt{Iodice17}).


The extinction corrected magnitudes and color derived are $m_{g}^{IGL} =
11.39 \pm 0.29$ mag, $m_{r}^{IGL}=10.61 \pm 0.07$ mag, and {\it (g -
  r)} $= 0.78 \pm 0.35$ mag,
and then the total luminosity in the {\it g} band is $L_{g}^{IGL} = 7.06 \times
10^{10} L_{\odot}$. The fraction of IGL, with respect to the total
luminosity of the group ( $L_{g} = 1.7 \times
10^{11} L_{\odot}$) is about 41\%, while with respect to the dominant
galaxy NGC 5018 ( $L_{g} = 1.5 \times
10^{11} L_{\odot}$) is about 47\%.
The integrated {\it (g - r)} color of the IGL component is consistent with those of the galaxies
in the group, within the errors.

Given that the IGL is composed of stars stripped from galaxies, its
color relative to the galaxies can give indications on the epoch when it
was stripped. In particular, if the IGL is redder than the
group galaxies, it is likely that stars have been stripped from the
galaxies at early times, while if the color of IGL is similar to that
of galaxies, it is likely that this component have formed from the
ongoing stripping of stars \citep{Krick06}.

The ongoing stripping scenario is consistent with our
results. Moreover, the color of the IGL component is also roughly
consistent, within the errors, with the color of the outskirts of NGC
5018. Simulations by \citet{Willman04} predict that $\sim$ 50\% of intragroup stars come
from bright galaxies, and as a consequence the color of the
intragroup stars should be consistent with the color of the outskirts
of bright group galaxies.

\section{Infrared analysis}\label{ir}

As shown in the previous sections, the photometric analysis in the
optical bands has shown the presence of a prominent and complex system
of dust lanes crossing the central regions of NGC 5018. Since the dust
optical depth decreases toward longer wavelengths, near-infrared (NIR)
photometry can help to reduce the dust absorption, that strongly
affects the starlight distribution in the galaxy. Moreover, adding
some infrared bands to the photometric analysis could help to better
understand the nature of the bridge connecting NGC 5018 to NGC 5022.

\begin{figure}
\centering
\hspace{-0.cm}
\includegraphics[width=7.5cm, angle=-0]{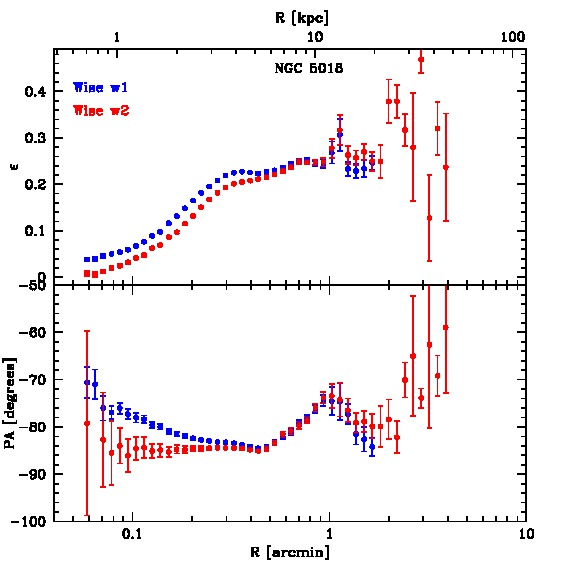}
\includegraphics[width=7.5cm, angle=-0]{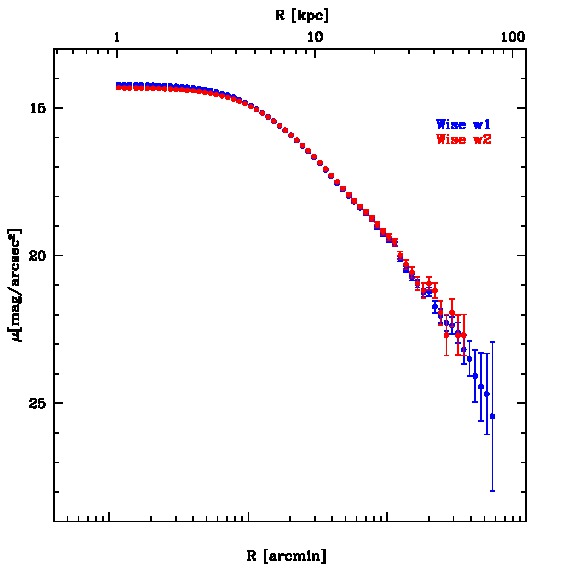}
\includegraphics[width=7.5cm, angle=-0]{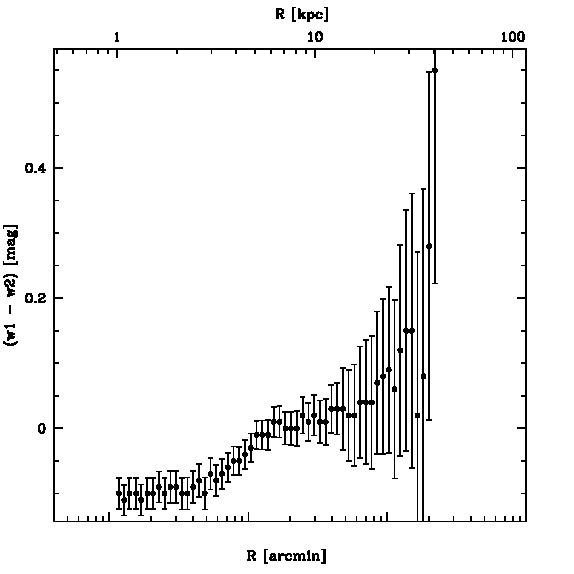}

\caption{{\it Top}: ellipticity
($\epsilon$) and position angle (P.A.) profiles for NGC 5018, in the Wise {\it w1}
(3.368 $\mu$m, blue points) and {\it w2} (4.618 $\mu$m, red points) bands. {\it Middle}: Azimuthally averaged
surface brightness profiles in the {\it w1}
(3.368 $\mu$m, blue points) and {\it w2} (4.618 $\mu$m, red points) bands. {\it Bottom}: Azimuthally averaged
color profile.}\label{wise}

\end{figure}

In this work we use images obtained with the Wide-field Infrared
Survey Explorer (WISE, \citealt{Wright10}) in the {\it w1} (3.368 $\mu$m) and
{\it w2} (4.618 $\mu$m) bands, to perform the infrared photometric
analysis for NGC 5018. The results of this analysis are shown in
Fig. \ref{wise}, where we plot the $\epsilon$ and $P.A.$ profiles (top
panel), the azimuthally averaged surface brightness profiles (middle
panel), and the mean {\it (w1 - w2)} color profile (bottom panel).

\begin{figure}
\centering
\hspace{-0.cm}
 \includegraphics[width=9cm, angle=0]{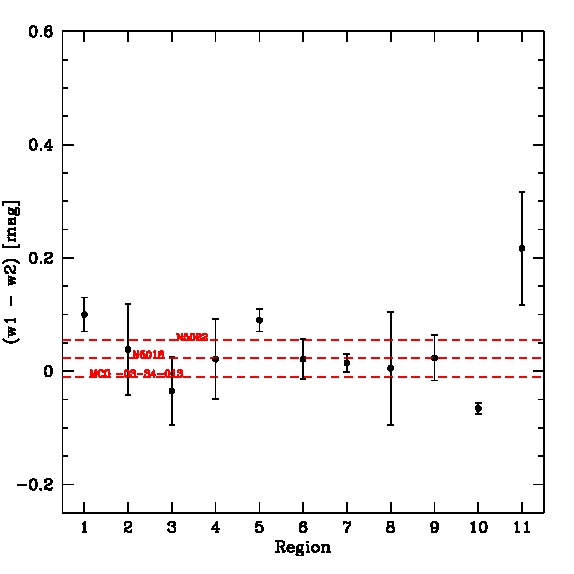}
\caption{The same as Fig. \ref{int_col} for the Wise {\it w1} band.}\label{int_col_wise}
\end{figure}

In order to better constrain the nature of the substructures observed in
the optical residual map of NGC 5018, we applied the same procedure
described in Sec. \ref{color} to the NIR residual map. We used the
same polygons obtained above (see Fig. \ref{res}), to estimate the integrated NIR color of
the substructure standing out form the map. We also derived the
integrated colors for NGC 5018 ({\it (w1-w2)}=0.02), NGC 5022 ({\it
  (w1-w2)}=0.05), and MCG-03-34-013 ({\it (w1-w2)}=-0.01), in order to
compare them with the colors of the different substructures.
The results are shown in
Fig. \ref{int_col_wise}, where we plot the infrared colors of the different substructures compared with
those of the three galaxies of the group.

This analysis confirms that also the NIR colors of the substructures
number 1  ({\it (w1-w2)}=0.1), number 5  ({\it (w1-w2)}=0.09), as well
as the bridge
(labelled as 10, ({\it (w1-w2)}=-0.06)) and the
loop (labelled as 11, ({\it (w1-w2)}=0.22)) are not consistent with any of the galaxies in
the group, as already found from the optical colors. 

\section{Ultraviolet analysis for NGC 5022}\label{UV}
In order to detect the presence of hot and young stars in the NGC 5018
group, we also performed a photometric analysis in the ultraviolet
(UV) bands. The near-ultraviolet (NUV) profile of NGC 5018 has been
published by \citet{Rampazzo07}, while the far-ultraviolet (FUV)
profile is not published because of the low S/N in the FUV band.

The FUV and NUV images of NGC 5022 used here, were obtained from GALEX archive (see
\citealt{Martin05,Morrissey05}). The exposure times were 4451 sec in NUV and 1568 sec in FUV bands (limiting magnitude in FUV/NUV $\sim$ 22.6/22.7 AB mag \citealt{Bianchi09}). 

We used FUV and NUV background-subtracted, intensity images derived
from the GALEX pipeline. FUV and NUV magnitudes have been computed as
m(AB)$_{UV}$ = -2.5 $\times$ log CR$_{UV}$ +ZP,
where CR is the dead-time-corrected, flat-fielded count rate and the zero points are ZP = 18.82 and ZP = 20.08 in the FUV and NUV respectively \citep{Morrissey07}. 

\begin{figure}
\centering
\hspace{-0.cm}
\includegraphics[width=9cm, angle=-0]{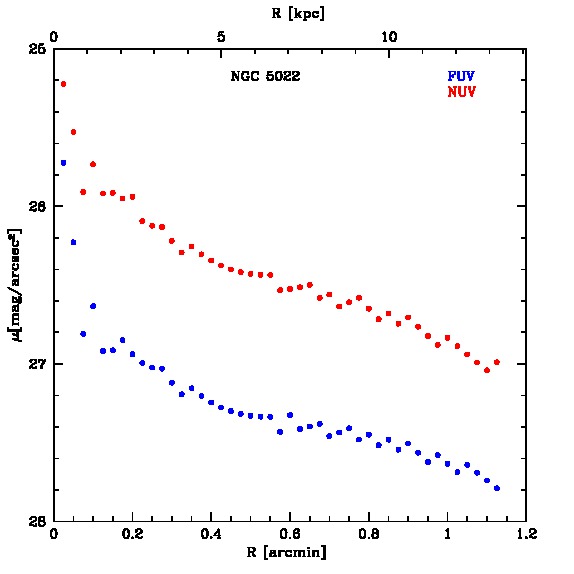}
\includegraphics[width=9cm, angle=-0]{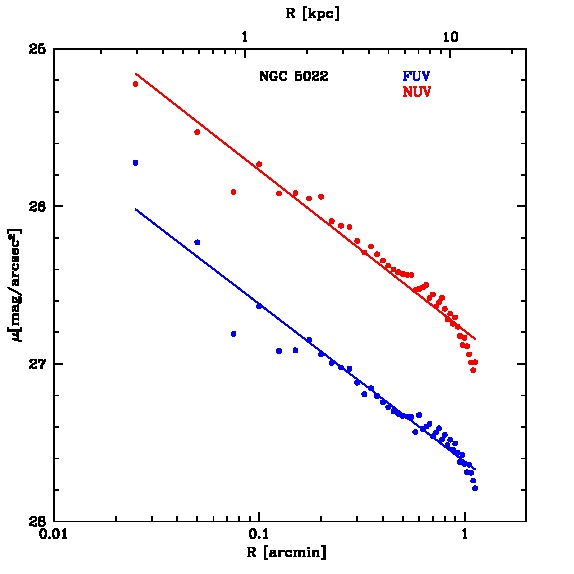}

\caption{{\it Top}: Azimuthally averaged
surface brightness profiles in the FUV
(blue points) and NUV (red points) bands. {\it Bottom}: The same
profiles plotted in logarithmic scale.}\label{prof_uv}

\end{figure}

Both FUV and NUV surface photometry have been performed using the IRAF
STSDAS ELLIPSE routine. The resulting SB profiles are shown in
Fig. \ref{prof_uv}. Our measured NUV and FUV total magnitudes (NUV=16.6 $\pm$ 0.03 and
FUV=17.4 $\pm$ 0.07) are consistent within errors with the one measured GALEX General Release 6. 
Both luminosity profiles are well fitted by a disk with n=0.9 (see the
bottom panel of Fig. \ref{prof_uv}).

\begin{figure*}
\centering
\hspace{-0.cm}
\includegraphics[width=18cm, angle=-0]{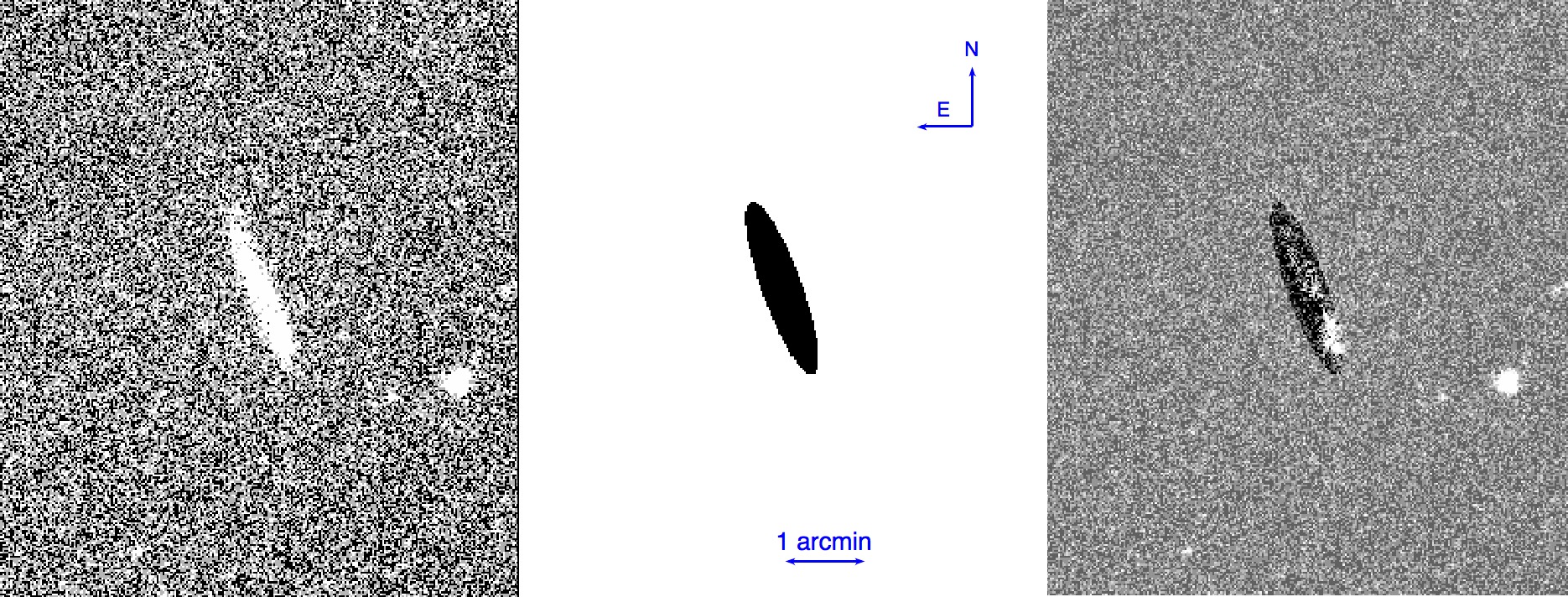}
\includegraphics[width=18cm, angle=-0]{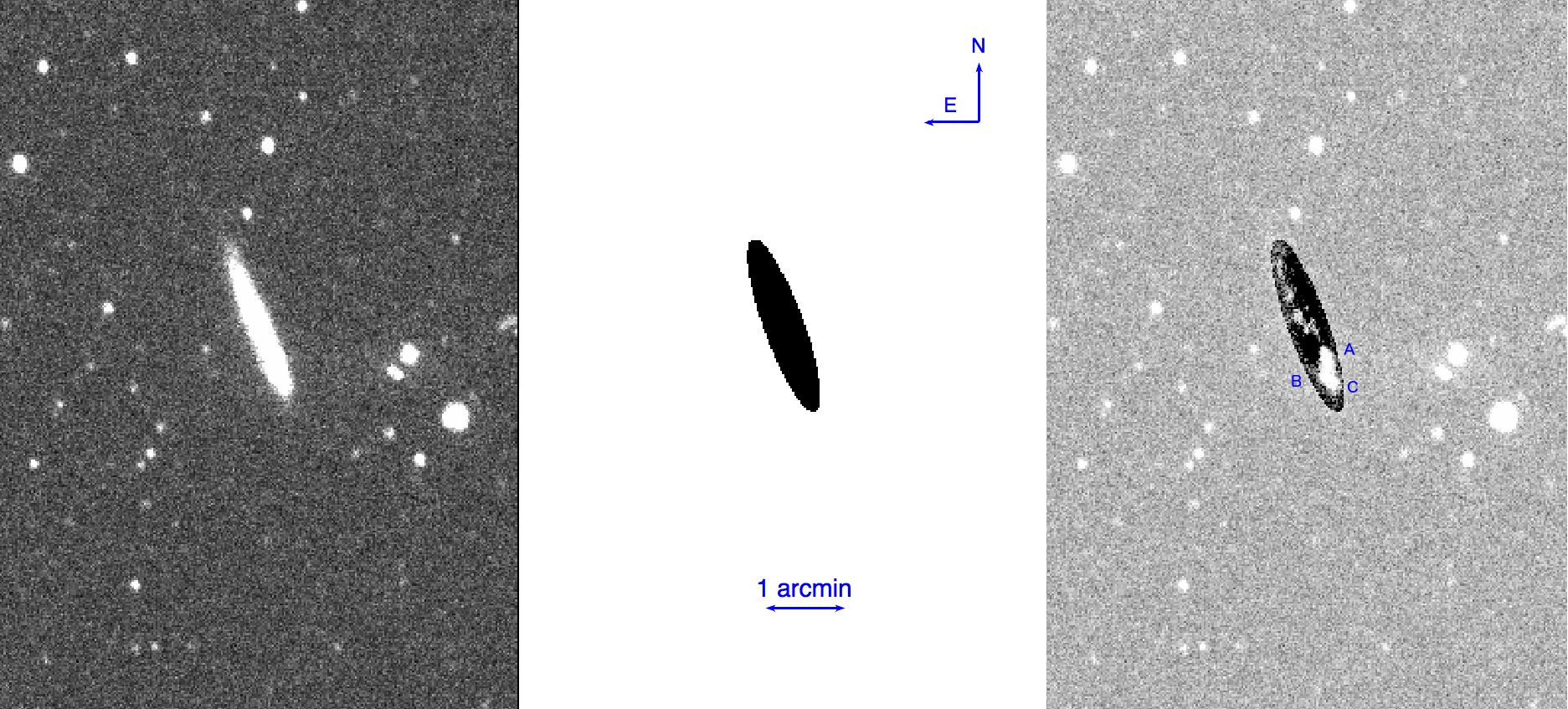}

\caption{{\it Top}: FUV image of NGC 5022 (left), 2D model of the
  galaxy (middle), residual of the subtraction of the model (right). {\it Bottom}: NUV image of NGC 5022 (left), 2D model of the
  galaxy (middle), residual of the subtraction of the model (right).}\label{mod_uv}

\end{figure*}

As for the optical images we used the IRAF task BMODEL to create a 2-D
image and then we subtract this model to the origin one. The residual
images are shown in Fig. \ref{mod_uv}. As we can see in the final
images (right panels) are well visible three bright regions in the southern part of the disk. We suspect these are region of star formation. For this reason we measured the flux of the three blobs visible. As a comparison we measured the total NUV and FUV fluxes for the whole north part of the galaxy.

Using the FUV flux we derived the present-day Star Formation Rate
(SFR) of each UV-bright blob (see the bottom-right panel of Fig. \ref{mod_uv}) following \citet{Kennicutt98}, using
its UV continuum luminosity and the relation SFR (M$_{\odot}$/yr) = 1.4 ×$\times$ 10$^{-28}$ L$_{FUV}$ (erg/s/Hz).
The final results are reported in Tab. \ref{tab:fos_indb}.
As we can see the SFR is the same in all the galaxy.

\begin{table*}
\caption{Integrated photometric UV properties}
\begin{tabular}{lcccccccccc}
\hline
Object & R.A. & DEC & {m$_{FUV}$(AB)} & {F$_{FUV}$} & {L$_{FUV}$}& {m$_{NUV}$(AB)} & {F$_{NUV}$} & {L$_{NUV}$} & {FUV-NUV} & {SFR$_{FUV}$}\\
 & (2000) & (2000) & &  & &  & & & & \\
\hline
\hline
Galaxy & 198.38  & -19.54 & 17.4$\pm$0.07 & 1.84 & 4.08 & 16.6$\pm$0.03 & 1.27 & 2.82 & 0.8 & 5.7\\
A  & 198.373  & -19.5578 & 18.42$\pm$0.19 & 0.75 & 1.64 & 17.64$\pm$0.21 & 0.65 & 1.44 & 0.92 & 2.32\\
B & 198.3748  & -19.5563 & 17.86$\pm$0.14 & 1.13 & 2.5 & 17.5$\pm$0.11 & 0.74 & 1.64 & 0.59 & 3.5\\
C  & 198.3741 & -19.5528 & 18.02$\pm$0.09 & 0.98 & 2.18 & 17.27$\pm$0.08 & 1.00 & 2.22 & 0.75 & 3.05\\

\hline
\end{tabular}\\
\label{tab:fos_indb}
\small{Notes: (1) Coordinates are given in degrees. 
(2) Fluxes are in units of $10^{-27}$ erg s$^{-1}$cm $^{-2}$$Hz^{-1}$.
(3) Luminosities are in units of $10^{26}$
erg s$^{-1}$ $Hz^{-1}$. (4)  SFRs are in units of $10^{-2}$ M$_{\odot}$ yr$^{-1}$}
\end{table*}

\section{Total accreted mass in a loose group of galaxies}\label{fit_sec}

Following the procedure adopted by \citet{Spavone17}, in order to
define the different components dominating the galaxies light
distribution at different scales,
we have described the surface
brightness profiles of NGC 5018 with a three component model: a
S{\'e}rsic profile \citep{Sersic63,Caon93} for the centrally concentrated in situ stars, a second
S{\'e}rsic for the ``relaxed'' accreted component, and another
S{\'e}rsic component for the diffuse and ``unrelaxed'' outer envelope
\citep{Seigar07,Donzelli11,Arnaboldi12,Cooper15,Iodice16,Spavone17}.

The stellar population of the central in situ component is expected to be similar to the dominant ``relaxed''
accreted component, while the outer diffuse component
representing ``unrelaxed'' accreted material (`streams' and other coherent
concentrations of debris) does not contribute significantly to the
brighter regions of the galaxy.

\begin{figure}
\centering
\hspace{-0.cm}
\includegraphics[width=8cm]{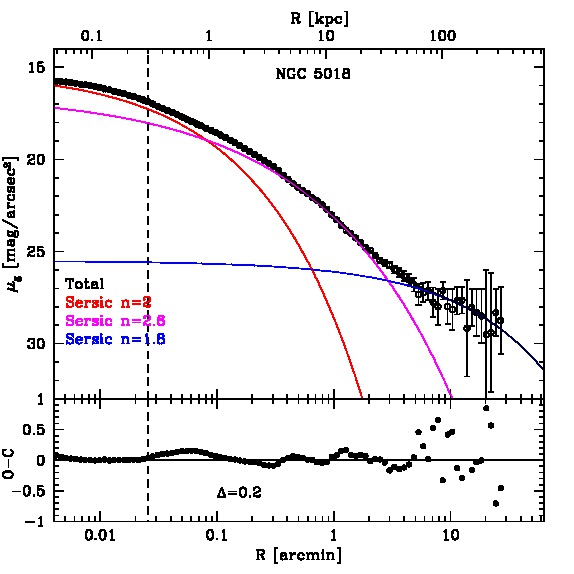}
 \includegraphics[width=8cm]{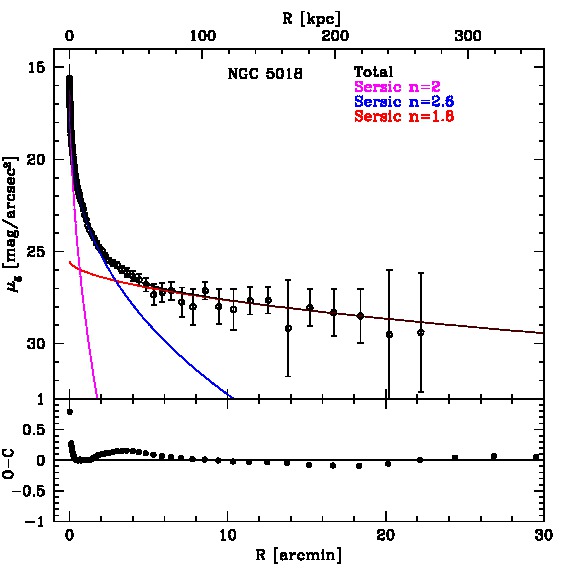}
\caption{VST {\it g} band profile of NGC 5018, in logarithmic (top)
  and linear (bottom) scale, fitted with a three component
  model motivated by the predictions of theoretical simulations (see
  \citet{Spavone17}).}\label{fit3comp}
\end{figure}

For the central ETG, NGC 5018, we used the same fitting procedure
described in \citet{Spavone17}, consisting in fixing $n \sim
2$ for the in situ component of our three-component fit, in order to mitigate the degeneracy
in parameters in such kind of fit. This typical value of $n$ has been
adopted on the basis of the results of \citet{Cooper13} for
massive galaxies. The results of this fit for NGC 5018, with the
corresponding (O-C) residuals, are shown in
Fig. \ref{fit3comp}, both in logarithmic (top panel) and in linear
(bottom panel) scale, while the best-fitting structural parameters for
the fit are reported in Tab. \ref{tabfit3comp}.
The result of the fit show that the outer part of this galaxy's
surface brightness profile is almost exponential in nature, and in
fact we checked that fitting an outer exponential component to the
surface brightness profile, does not change the residuals and the
shape of the total fitted profile.

\begin{figure}
\centering
\hspace{-0.cm}
 \includegraphics[width=8cm, angle=-0]{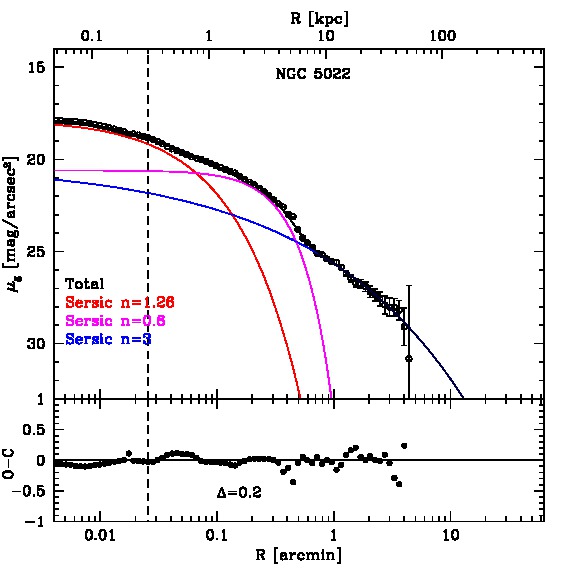}
 \includegraphics[width=8cm, angle=-0]{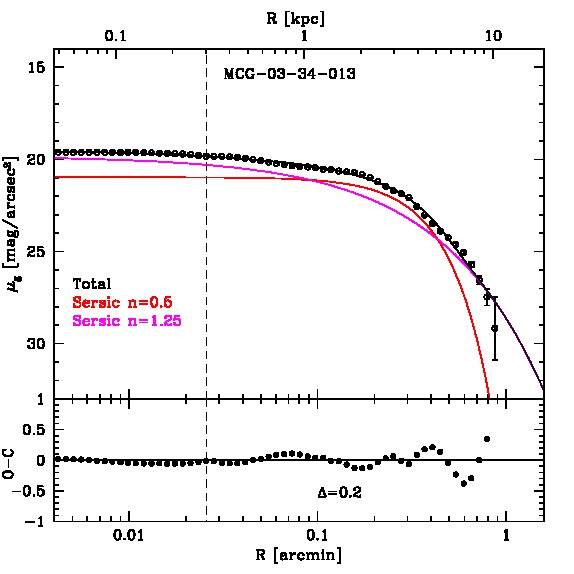}

\caption{VST {\it g} band profiles of NGC 5022 (top), fitted with a three components
  model, and MCG-03-34-013 (bottom), fitted with a double Sersic model.}\label{fit}

\end{figure}

The surface brightness profiles of NGC 5022 and MCG-03-34-013 have
also been fitted with a three and two component model, respectively, in which all the
parameter were left free. The results of the fit are shown in
Fig. \ref{fit} and the best-fitting parameters are reported in Tab. \ref{tabfit3comp}.


\begin{table*}
\setlength{\tabcolsep}{1.0pt}
\scriptsize
\caption{Best-fitting structural parameters for a three-component fit.} \label{tabfit3comp}
\vspace{10pt}
\begin{tabular}{lccccccccccccccccc}
\hline\hline
Object & $\mu_{e1}$ &$r_{e1}$&$n_{1}$& $\mu_{e2}$ &$r_{e2}$&$n_{2}$&$\mu_{e3}$
&$r_{e3}$&$n_{3}$&$m_{T,1}$&$m_{T,2}$&$m_{T,3}$& $f_{h,T}$\\ 
    & [mag/arcsec$^{2}$] &[arcsec]& & [mag/arcsec$^{2}$] & [arcsec]&&
    [mag/arcsec$^{2}$] & [arcsec]&[mag]&[mag]&[mag]& \\ 
\hline \vspace{-7pt}\\
NGC 5018  & 19.10$\pm$0.10 & 5.24$\pm$0.07  & 2&21.67
$\pm$0.34&31.75$\pm$0.50&2.6$\pm$0.1& 29.06$\pm$0.20 &1502$\pm$7
&1.8$\pm$0.2&12.16 &10.77&9.79&92\%\\ 
 NGC 5022    & 20.15$\pm$0.50 & 3.00$\pm$0.20  & 1.26$\pm$0.35&21.55
$\pm$1.27&12.27$\pm$2.90&0.60$\pm$0.20& 26.38$\pm$0.32 &87.61$\pm$7.6
&3.00$\pm$0.62&14.38&12.72&13.28&88\%\\ 
 MCG-03-34-013 & 21.70$\pm$0.06 & 12.09$\pm$2.00  & 0.50$\pm$0.34&22.18
$\pm$0.18&11.55$\pm$0.02&1.25$\pm$0.03& - &-
&-&12.16&10.77&-&78\%\\ 
\hline
\end{tabular}

\tablecomments{Columns 2, 3, and 4 report effective magnitude and effective
radius for the inner component of each fit. The S{\'e}rsic index for
the in situ component of the BCG NGC 5018 was fixed to $n\sim 2$ using the models as a
prior \citep{Cooper13}, as explained in \citet{Spavone17}.  We allowed small variations of $\pm 0.5$ around the mean
  value of $n=2$. This would bracket the range of $n$ in the
  simulations and allows us to obtain a better fit. Columns
5, 6, 7, 8, 9 and 10 list the same parameters for the second and the
third components. Columns 11, 12, and 13 report the
total magnitude of the inner ($m_{T ,1}$) and outer S{\'e}rsic
components ($m_{T ,2}$ and $m_{T ,3}$). Column 14 gives the
total accreted mass fraction derived from our fit.}
\end{table*}

We used these fits to derive the total magnitude of the different S{\'e}rsic
components, $m_{T,1}$, $m_{T,2}$, and $m_{T,3}$, as well as the
relative contribution of the accreted component with respect
to the total galaxy light, $f_{h,T}$, reported in
Tab. \ref{tabfit3comp}. The total accreted mass fraction for the
galaxies in this small group, ranges between 78\% and 92\%. These
values are consistent with the results of numerical simulations,
which predict that stars accreted by BCGs account for most of
the total galaxy stellar mass \citep{Cooper13,Cooper15,Pillepich18}.

These results can be used to estimate the stellar mass fractions in
the different components by assuming appropriate stellar mass-to-light
ratios, in order to compare the accreted mass ratios we infer from
our observations with other observational
estimates for BCGs of similar total mass by \citet{Seigar07,
  Bender15,Iodice16,Spavone17}, and with theoretical predictions from semi-analytic particle-tagging
simulations by \citet{Cooper13,Cooper15}, and the Illustris cosmological
hydrodynamical simulations \citep{Pillepich18}.

\begin{figure}
\centering
\hspace{-0.cm}
 \includegraphics[width=9cm, angle=-0]{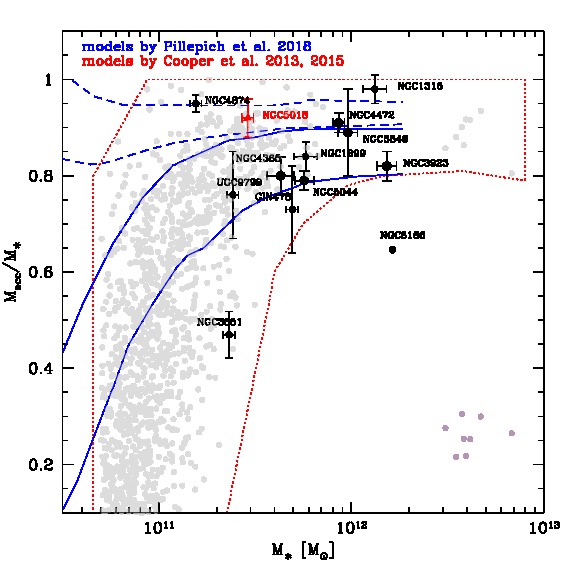}
\caption{Accreted mass fraction vs. total stellar mass for ETGs. The
 measurement for NGC 5018 is given as red triangle. Black circles correspond to other BCGs from the literature
 \citep{Seigar07, Bender15, Iodice16, Spavone17}. Red region
   indicates the
 predictions of cosmological galaxy formation simulations by
 \citet{Cooper13,Cooper15}. Blue continuous and dashed regions
 indicate the accreted mass fraction measured within 30 kpc and
 outside 100 kpc, respectively, in Illustris simulations by
 \citealt{Pillepich18} (see their Fig. 12).
 Purple-grey points show the mass fraction associated with the
 streams from Tab. 1 in \citet{Cooper15}.}\label{halo}
\end{figure}

To this aim, we measured the mean {\it (u -
    g)} and {\it (g - r)} colors for each galaxy in regions where the
  central galaxies and the outer envelopes dominate, obtaining the
  values reported in Tab. \ref{mass}. We then used stellar population synthesis models \citep{Vazdekis12,
Ricciardelli12}, with a Kroupa IMF, to derive the mass-to-light ratios
corresponding to the average colors, and hence the stellar mass of the
whole galaxy and of the outer envelope.
These results are summarized in Table \ref{mass} and the comparison is
shown in Fig. \ref{halo}. Since simulations used to compare the
  stellar halo mass fraction of our galaxies cover scales from the
  stellar haloes of Milky Way-like galaxies to the cD envelopes of
  groups and clusters. For this reason the comparison shown in the
  Fig. \ref{halo} is meaningful only for the more massive galaxy of the group.

\begin{table*}
\setlength{\tabcolsep}{2.5pt}
\begin{center}
  \caption{Total and accreted stellar masses of galaxies in our sample.} \label{mass}
\vspace{10pt}
\begin{tabular}{lcccccccccccccccc}
\hline\hline
Object & $({\it u-g})$ &$({\it g-r})$&$(M/L)_{g}$&$M^{*}_{tot}$&$M^{*}_{total\ accreted}$\\
    & [mag] &[mag]&[$M_{\odot}/L_{\odot}$] &[$M_{\odot}$]&[$M_{\odot}$]\\
\hline \vspace{-7pt}\\
 NGC 5018    &    1.6 $\pm$ 0.4   &   0.7 $\pm$ 0.2  & 1.97& $2.9\times10^{11}$&$2.7\times10^{11}$\\
NGC 5022    &    1.0 $\pm$ 0.4   & 1.6 $\pm$ 0.5  &  1.48 & $1.6\times10^{10}$ &$1.4\times10^{10}$\\
MCG-03-34-013    &  0.6 $\pm$ 0.1  & 1.4 $\pm$ 0.1  & 0.82 &$4.3\times10^{9}$ &$3.3\times10^{9}$\\
\hline
\end{tabular}
\tablecomments{Columns 2, 3 and 4 show the mean, extinction corrected, {\it
    u - g} and {\it g-r} colours of each galaxy and the relative mass-to-light
  ratios in the {\it g} band. Columns 5 is the galaxy stellar mass,
  while column 6 reports the total accreted stellar masses, derived from the
  three and two-component fit.} 
\end{center}
\end{table*}

We find that the stellar
mass fraction of the accreted component derived for the three main galaxies in the
NGC 5018 group is
fully consistent both with published data for other BCGs
\citep{Seigar07, Bender15, Iodice16, Spavone17}, despite
considerable differences in the techniques and assumptions involved, and with
the theoretical models by \citet{Cooper13,Cooper15} and by \citet{Pillepich18}. 

\section{Discussion and Conclusions}\label{conc}



In this work we present a new deep mosaic of $1.2 \times 1.0$ square degrees of the group of galaxies centered on NGC~5018, acquired at the ESO VLT Survey Telescope.
Taking advantage of the deep and multiband photometry (in {\it u, g, r} bands) and of the large field of view of the
VST telescope, we studied the structure of the galaxy members and the faint feature into the intra-group space and, then, we give an estimate of the intragroup diffuse light in the NGC 5018 group of galaxies.

\subsection{Tracing the build-up history of the group}

NGC 5018, sometimes classified as E3, and at South-East NGC 5022
        which is  an edge-on spiral, seem to be linked by a
        very low surface brightness filament. Indeed, many sources in
        the literature identify these two as an interacting pair of
        galaxies. 

The spiral at South-East shows no warp or obvious signs of
interaction, while N5018 exhibits a perturbed morphology in the outskirts, with shells and ripples, as well as
 inner dust signs. 
 The  HI map from the VLA (Fig. \ref{field}) indicates that there is a long tail of gas that 
connects NGC 5022 to the another group member, MCG-03-34-013, on the North-West with respect to NGC~5018. 
Such a HI filament is on the north and does not crosses NGC 5018, although a branch of it does  enter the galaxy.
Therefore, one open issue in the formation history of the group is if the tails and all intra-group features visible in the light distribution are
tracing the same interaction traced by the cold gas filaments.

Taking advantage by the multi-band observations from VEGAS, we are able to estimate the integrated colors of all the 
intra-group features (see Sec.~\ref{icl}). We found that 
the integrated optical colors of all substructures (see Fig. \ref{res}), are not consistent with the average colors of any of the
galaxies in the group, being redder. This is confirmed by estimating the NIR integrated colors for the same regions and galaxies.
Therefore, this finding suggests an external  origin for them.

Something similar was found for the bright galaxy NGC~1316, in SW subgroup of the Fornax cluster \citep{Iodice17a}, where, 
by adopting the same approach based on the color analysis, authors concluded that some of the substructures in the envelope 
come from a recent accretion event of smaller and bluer galaxy, while large and redder tidal tail are more reasonably related to an earlier
interaction event.

In NGC~5018, the analysis of the integrated colors suggests that the plume at the NW side and all the substructures in its central
spheroid, are coeval with the galaxy, given that they have comparable colors. These features can thus be interpreted as material expelled
from NGC 5018 during a merging event in its formation history.
Differently, the bridge between NGC~5018 and NGC 5022, the diffuse loop on the est side of NGC 5022, and the
patches numbered as 1 and 5 on East and SW side (see Fig.\ref{res}), have both optical and NIR colors not consistent with the
average colors of any of the galaxies in the group, which are in the range $g-r\sim0.6-0.75$~mag, being redder ($g-r\sim 0.9 -1.1$~mag, see Fig.\ref{int_col}).
They are more similar to the colors in the outskirts of NGC~5018 and NGC 5022, which are $g-r\sim0.9-1.4$~mag for $1\leq R \leq 4$~arcmin (see Fig.\ref{col_5018}). Therefore, we claim that such a features could be made by a stripped material from the galaxy outskirt in a close passage.
In particular, the almost polar half-ring on the East side of NGC~5022
could formed by the interaction with an high inclined orbit \citep[see
e.g.][]{Bournaud03}. Moreover, such an interaction might have
triggered the new star formation regions in the south part of
NGC~5022, as found from the analysis of the UV data (see Sec. \ref{UV}).

On the other hand, the long tail of HI gas is tracing the interaction between NGC 5022 and MCG-03-34-013  \citep[see also][]{Kim88,Guhathakurta90}.

\subsection{Total accreted mass and intra-group light estimates}

According to \citet{Spavone17}, by fitting the SB profile of NGC
5018 with three S{\'e}rsic components, we were able to have an estimate of the total accreted mass in this galaxy, 
which is the brightest group member, and for all the less brighter galaxies (see Sec\ref{fit_sec}).
We found that the total accreted mass fraction for the
galaxies in this small group, ranges between 78\% and 92\%. These
values are consistent with the results of numerical simulations by \citet{Cooper13,Cooper15,Pillepich18} 
and with published data for other BCGs \citep{Seigar07, Bender15, Iodice16, Spavone17}, despite
considerable differences in the techniques and assumptions involved (see Fig.\ref{halo}).


The change in the trend of the ellipticity and P.A. with radius appears to
correlate with an inflection in the SB profile of NGC 5018, which also
marks the transition between the two accreted components in the fit of
the light distribution (see Fig. \ref{fit3comp}). As already pointed
out in \citet{Spavone17}, such inflection could mark the transition
between two components in different states of dynamical relaxation. 
The upward inflection of the profile over the transition radius,
occurs well beyond 2$R_{e}$, suggesting an excess of weakly-bound
stars associated with a recent accretion event. NGC 5018 shows clear
signs of ongoing interaction and  accretion events, indicating
that its outer regions are still being assembled, consistent with theoretical expectations for such galaxies (e.g. \citealt{Cooper15}).

 The total {\it g}-band luminosity of the IGL is $L_{g}^{IGL} = 7.06 \times
10^{10} L_{\odot}$, which is $\sim$  41\% of the total luminosity of the group. The
    IGL has {\it (g - r)} $= 0.78 \pm\ 0.35$ mag, which is similar to the color in the
    halo of the BCG in the group core, NGC 5018, and consistent with
    the integrated colors of  the other galaxies in the group.
 This is consistent with that found in many compact group of galaxies
\citep{DaRocha05,White03,DaRocha08}, and with the prediction of
numerical simulations \citep{Sommer06}. The presence of IGL indicates
that tidal encounters stripped a considerable amount of mass from the
member galaxies, and that the group is in an advanced stage of its
dynamical evolution \citep{DaRocha05,Sommer06,Rudick06}.
Slow encounters in group environment are in fact effective in liberating stars in the
intragroup medium, and since the fraction of IGL increases with the
degree of dynamical evolution of the group/cluster, it can be used as
a ``dynamical clock'': more evolved groups/clusters have largest
fractions of diffuse light.

Moreover, as already argued by \citet{Krick07}, the color of IGL
and its correlation with the color of galaxies in the group, could
help to constrain the epoch at which stars forming the IGL have been stripped.
In the case of NGC 5018, since the color of the IGL component is
consistent with those of galaxies belonging to the group, the ongoing
stripping scenario is the most plausible for the formation of
intragroup light in this system.

As a conclusive remark,  the picture emerging from the
multi-wavelength study illustrated in this work is that 
there are at least two different interactions involving the group members: one between the two brightest galaxies NGC~5018 and NGC~5022, which generates the tails and ring-like structures detected in the light, and another between the two gas-rich galaxies of the group, NGC~5022 and MCG-03-34-013 that have produced the long HI tail connecting the two systems.
Moreover, a minor merging event also happened in the formation history of NGC~5018 that have perturbed the inner structure of this galaxy.

The unperturbed isophotes of MCG-03-34-013 would suggest that the interaction involving such a small galaxy  with NGC~5022 is more likely 
an high-velocity encounter between them, while the gravitational forces between the two bright group members were strong enough to perturb the outskirts and generate intra-group material.



\acknowledgements
We are very grateful to the anonymous referee for his/her comments and suggestions which helped us to improve and clarify our work.
This work is based on visitor mode observations taken at the ESO
    La Silla Paranal Observatory within the VST Guaranteed Time
    Observations, Programme IDs 096.B-0582(B), 097.B-0806(A) and
099.B-0560(A). MS wishes to thank the ESO staff of the
    Paranal Observatory for their support during the observations at
    VST. MS acknowledge finacial support from the VST project
      (P.I. P. Schipani). R.R. and D.B acknowledge funding from the INAF PRIN- SKA 2017 program 1.05.01.88.04. This publication makes use of data products
      from the Wide-field Infrared Survey Explorer, which is a joint
      project of the University of California, Los Angeles, and the
      Jet Propulsion Laboratory/California Institute of Technology, funded by the National Aeronautics and Space Administration.



\end{document}